\documentclass[iop]{emulateapj}

\usepackage{paralist}
\usepackage{graphics,graphicx}
\newcommand{\nar}{New Astronomy Reviews}

\begin{document}

\shortauthors{Mineo et al.}
\shorttitle{The radial distribution of X-ray Binaries and Globular Clusters in NGC~4649}
\title{The radial distribution of X-ray Binaries and Globular Clusters in NGC~4649 and their relation with the local stellar mass density}

\author{S. Mineo\altaffilmark{1,2}, G. Fabbiano\altaffilmark{1}, R. D'Abrusco\altaffilmark{1}, T. Fragos\altaffilmark{1}, D.-W. Kim\altaffilmark{1}, J. Strader\altaffilmark{3}, J. P. Brodie\altaffilmark{4}, J. S. Gallagher\altaffilmark{5}, A. Zezas\altaffilmark{6}, B. Luo\altaffilmark{7}}

\altaffiltext{1}{Harvard-Smithsonian Center for Astrophysics, 60 Garden Street Cambridge, MA 02138 USA; smineo@head.cfa.harvard.edu} 
\altaffiltext{2}{Department of Physics, University of Durham, South Road, Durham DH1 3LE, UK}
\altaffiltext{3}{Department of Physics and Astronomy, Michigan State University, East Lansing, Michigan 48824, USA} 
\altaffiltext{4}{UCO/Lick Observatory, 1156 High St., Santa Cruz, CA 95064, USA}
\altaffiltext{5}{Department of Astronomy, University of Wisconsin, Madison, WI 53706-1582, USA}
\altaffiltext{6}{Physics Department, University of Crete, P.O. Box 2208, GR-710 03, Heraklion, Crete, Greece}
\altaffiltext{7}{Department of Astronomy \& Astrophysics, 525 Davey Lab, The Pennsylvania State University, University Park, PA 16802, USA}

\begin{abstract}
We investigate the radial distribution of the low-mass X-ray binary (LMXB) population in the elliptical galaxy NGC~4649, using {\it Chandra} and {\it Hubble} data to separate the field and globular cluster (GC) populations. GCs with LMXBs have the same radial distribution as the parent red and blue GCs. The radial profile of field LMXBs follows the $V$-band profile within the $D25$ of NGC~4649. Using the spatial information provided by our data, we find that the global galaxy-wide relations between cumulative number and luminosity of LMXBs and the integrated stellar mass hold on {\em local} scales within $D25$. An excess of field LMXBs with respect to the $V$-light is observed in the galaxy's outskirt, which may be partially due to unidentified GC sources or to a rejuvenated field LMXB population caused by past merging interactions.
\end{abstract}


\section{Introduction}
\label{sec:intro}

Given the old age of the stellar populations of elliptical galaxies, there is no doubt that the X-ray emitting accretion binaries detected in copious numbers in these galaxies with {\it Chandra} \citep[see review,][]{2006ARA&A..44..323F} are low-mass X-ray binaries (LMXBs), with sub-solar companion stars.  These LMXB samples have rekindled the discussion on the origin of LMXBs that began with their discovery in the Milky Way \citep[see][]{1974IAUS...64..147G}. {\it Chandra} observations show that LMXBs are found both in globular clusters (GCs) and in the stellar field. Dynamical formation in GCs is the most efficient way to form LMXBs, which could then disperse in the field because of formation kicks or evaporation of the parent cluster \citep{1975ApJ...199L.143C,1984ApJ...282L..13G, 2002ApJ...574L...5K, 2004ApJ...607L.119B, 2007ApJ...662..525K}. While less efficient, the evolution of a native binary in the stellar field is also a viable LMXB formation channel (Verbunt \& van den Heuvel 1995). Both mechanisms are likely to take place in elliptical galaxies, since their LMXB content  depends on both the total stellar mass of the galaxy and its GC specific frequency \citep{2004ApJ...611..846K, 2004MNRAS.349..146G}. Moreover,  at the lower luminosities ($L_{\rm{X}} < 10^{37}\,\rm{erg}\,\rm{s}^{-1}$), the X-ray Luminosity Functions (XLF) of GC and field LMXB clearly differ, suggesting different families of sources \citep{2007MNRAS.380.1685V, 2009ApJ...701..471V, 2009ApJ...703..829K, 2011A&A...533A..33Z}. 

Not all GCs are similarly effective at forming LMXBs. Several factors may affect the probability of dynamical LMXB formation. A factor is the stellar encounter rate, which is a function of the GC mass and compactness/radius \citep{2006csxs.book..341V, 2007ApJ...660.1246S, 2007ApJ...671L.117J, 2010MNRAS.407.2611P}. The galacto-centric radius has also been noted, in the sense that  more centrally located GCs have been reported in some cases to be more likely to host LMXBs; however, this also may be a byproduct of the stellar encounter rate, since centrally-located GCs may also be more compact \citep{1962PASP...74..248H, 1991ApJ...375..594V, 2006ARA&A..44..193B, 2010ApJ...715.1419M, 2010ApJ...725.1824F}. Finally, most GC-LMXBs reside within red, metal-rich GCs \citep{1993AdSpR..13..597G, 1995ApJ...439..687B, 2007ApJ...662..525K, 2002ApJ...574L...5K, 2004ApJ...606..430M, 2004ApJ...613..279J, 2006ApJ...647..276K, 2007ApJ...660.1246S}. The metallicity effect is the strongest, as demonstrated by a recent study by \citet{2013ApJ...764...98K} making use of all the available {\it Chandra} and {\it Hubble} data, which also finds that the ratio of GC-LMXB fractions in metal-rich to metal-poor GCs is $3.4\pm0.5$ for a wide range of X-ray luminosities. \citet{2012ApJ...760L..24I} suggest that these differences can be explained by the increase in number densities and average masses of red giants in higher metallicity populations, where red-giants operate as seeds for LMXB dynamical formation.

While the bulk of the above conclusions on LMXB formation in elliptical galaxies has been reached based on the study of the XLFs (and global content of LMXBs), the spatial distribution of LMXBs may offer different important clues. The results so far are contradictory. In particular, some authors concluded that overall the LMXB distribution follow closely the stellar light \citep{2003ApJ...586..826K, 2004ApJ...612..848H, 2006ApJ...647..276K}, and comparisons of the radial distributions of field- and GC-LMXBs in a few galaxies did not reveal significant differences \citep{2003ApJ...595..743S, 2004ApJ...613..279J, 2006ApJ...647..276K}, but these studies may have been affected by poor statistics as well as by heterogeneous selection of GCs from {\em Hubble Space Telescope (HST)} and ground-based (KPNO 4m) observations. \citet{2007ApJ...662..525K} instead concluded that GC and field LMXBs follow different distributions, therefore suggesting that field LMXBs should be formed in the stellar field. Based on overall galaxy properties, \citet{2005ApJ...631..511I} and \citet{2005ApJ...621L..25J} argued for in situ formation of field LMXBs, while \citet{2009ApJ...703..829K} suggested the possibility of a mixed origin \citep[see also][]{2005ApJ...631..511I}.

Our large coverage with both {\it Chandra} and {\it HST} of the Virgo elliptical NGC~4649 \citep{2012ApJ...760...87S, 2013ApJS..204...14L}, enables us to revisit this question, with a rich population of several hundreds LMXBs, and a homogeneous and complete GC population. 
The present work is based on the X-ray source catalog of the elliptical galaxy NGC~4649 by \citet{2013ApJS..204...14L}, and on the associated optical catalog of globular clusters detected with {\em HST} by \citet{2012ApJ...760...87S}. We adopt a distance to NGC~4649 of 16.5 Mpc. 
In Section~\ref{sec:data} we discuss our data samples; in Section~\ref{sec:lmxb_spatial} we analyze the radial distribution of GCs (both red and blue) and their associated LMXBs; we  derive the radial distributions of field LMXBs and compare it with that of GC-LMXBs, GCs and stellar light, which is a proxy of the integrated stellar mass. In Section~\ref{sec:disc} we discuss our results. Our main conclusions are summarized in Section~\ref{sec:soncl}.

\section{Data}
\label{sec:data}

\begin{figure}
\begin{center}
\includegraphics[width=1\linewidth]{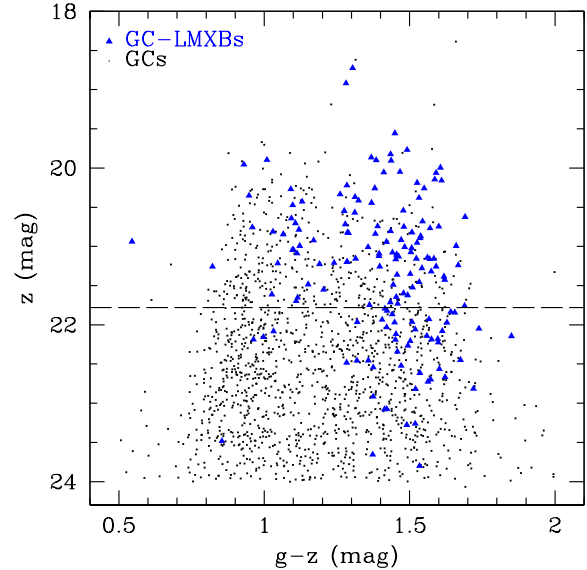}
\caption{$z$ vs. $g-z$ color-magnitude diagram of GCs in NGC 4649. The black points are all GCs from \citet{2012ApJ...760...87S} sample (1603 sources). The blue triangles are those (157) matched with the X-ray sources from \citet{2013ApJS..204...14L} catalog (see Sect.~\ref{sec:gc_lmxb_spatial} for details). The horizontal dashed line indicates the $> 90\%$ completeness level for ground-based data (see Sect. \ref{sec:field_lmxb_spatial} for details), for comparison with {\it HST} data.}
\label{fig:color_mag}
\end{center}
\end{figure}

The data are from our full-coverage joint {\it Chandra}-{\it HST} survey of NGC~4649 (P.I. Fabbiano). The original samples comprise 1603 GCs \citep{2012ApJ...760...87S} and 501 X-ray sources \citep{2013ApJS..204...14L}. From these samples we excluded all sources detected within the $D25$ ellipse of the nearby spiral galaxy NGC~4647 and those within the central ($R< 0.17\arcmin$) region of NGC~4649; the latter to avoid confusion. We thus obtained 1516 GCs, of which 731 are blue, $(g-z)<1.18$, and 785 red, $(g-z) \gtrsim 1.18$ ($g$ and $z$ are both AB magnitudes). These GCs and the associated 157 X-ray sources are plotted in the $z$ vs $(g-z)$ space in Fig.~\ref{fig:color_mag}. As expected \citep[see][and references therein]{2006ARA&A..44..323F}, most X-ray sources are hosted by red and luminous GCs. The samples used in this work are summarized in Table~\ref{table:data}.

There are 280 X-ray sources with no GCs counterpart, of which 74 reside outside the {\it HST} survey area. To establish if some of these sources are associated with GCs, we matched their positions with the GCs in the catalog of \citet{2008ApJ...674..857L}. The latter, from KPNO 4m telescope images, reaches to a projected radius of $\sim$$8.2\arcmin$ and is $> 90\%$ complete to $T_{1}\sim23$ ($V\sim22.5$). We excluded the resulting 12 GC matches from the field-LMXBs sub-sample and converted their $(C-T_{1})$ colors into $(g-z)$ in order to separate blue and red GCs. All of them are associated with luminous GC candidates, for which contamination effects are low. For comparison with {\it HST} data, we converted the value of $V$ mentioned above into $z$, using the equation $V = 0.753\times (g-z) - 0.108 + z$ (Usher et al. 2013, ApJ submitted), assuming an average $(g-z) =1.1$. We show the $> 90\%$ completeness level for ground-based data in Fig.~\ref{fig:color_mag}, which suggests that a fraction of low luminosity GCs, and therefore their GC-LMXBs may still remain undetected in the ground-based data.

For uniformity and sensitivity, only GCs and GC-LMXBs from the {\em HST} sample were used for the data analysis throughout the paper.
Using the $z$-band completeness curves from \citet{2009ApJS..180...54J} for the typical GC half-light radius $r=0.0385\arcsec$, the GC sample is complete for $z<22.2\,\rm{mag}$ at all radial bins of our grid. However, when comparing two distributions equally affected by incompleteness, we include faint GCs to maximize statistics; these cases will be specified at their occurrence.

We adopted a common threshold luminosity for the X-ray sources, at $4.8\times 10^{37}\,\rm{erg}\,\rm{s}^{-1}$, corresponding to $50\%$ completeness at the $11^{\rm{th}}$ radial bin ($5.0\arcmin$ to $5.5\arcmin$ from the galaxy center, enclosing the $D25$ ellipse). We corrected the radial distribution of field LMXBs using the spatially-resolved completeness function $K(L)$ of \citet{2013ApJS..204...14L}. Throughout this paper we account for the contribution of background active galactic nuclei (AGNs) to the X-ray source counts in the field, both in terms of number and luminosity, using the $\log N - \log S$ determined by \citet{2008MNRAS.388.1205G}. This correction is not needed for GC-LMXBs, thanks to the matches with GCs.

\begin{table*}
\tablewidth{0pt}
\tabletypesize{\scriptsize}
\centering
\begin{minipage}{100mm}
\caption{Summary of the data}
\label{table:data}
\begin{tabular}{@{}l c c c l c c c@{}}
\hline
\hline
\vspace{1mm}
& \multicolumn{3}{c}{{\sc Globular Clusters}} &  \multicolumn{4}{c}{{\sc X-ray sources}}\\
\,\,\,&\vline\, red & blue & tot & \vline\, red GCs & blue GCs & field & tot \\
\hline
All &\vline\, 785 & 731 & 1516 & \vline\, 128 & 29 & 268 & 425 \\
Complete &\vline\, 329 & 297 & 626 & \vline\, 59 & 18 & 151 & 228 \\
AGN-subtracted &\vline\, - & - & - & \vline\, 59 & 18 & 120 & 197 \\
\hline
\vspace{0.1mm}
\end{tabular}
Note. Blue: $(g-z)<1.18$, red: $(g-z) \gtrsim 1.18$, X-ray completeness limit $L_{\rm{lim}}=4.8\times 10^{37}\,\rm{erg}\,\rm{s}^{-1}$, optical GC completeness limit: $z<22.2\,\rm{mag}$.
\medskip
\end{minipage}
\end{table*}

\begin{figure*}
\begin{center}
\begin{minipage}{160mm}
\hbox
{
\includegraphics[width=77mm]{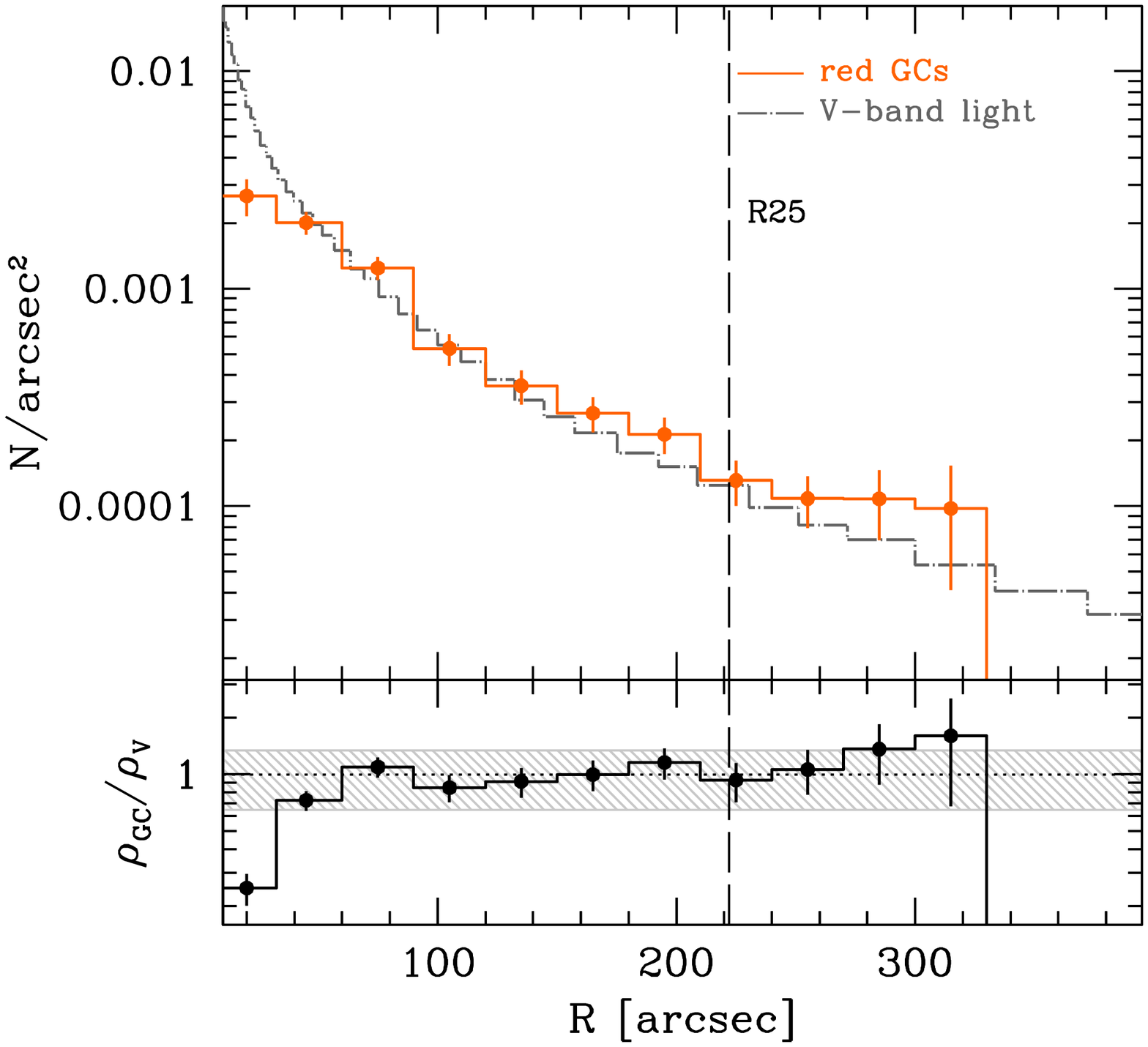}
\includegraphics[width=77mm]{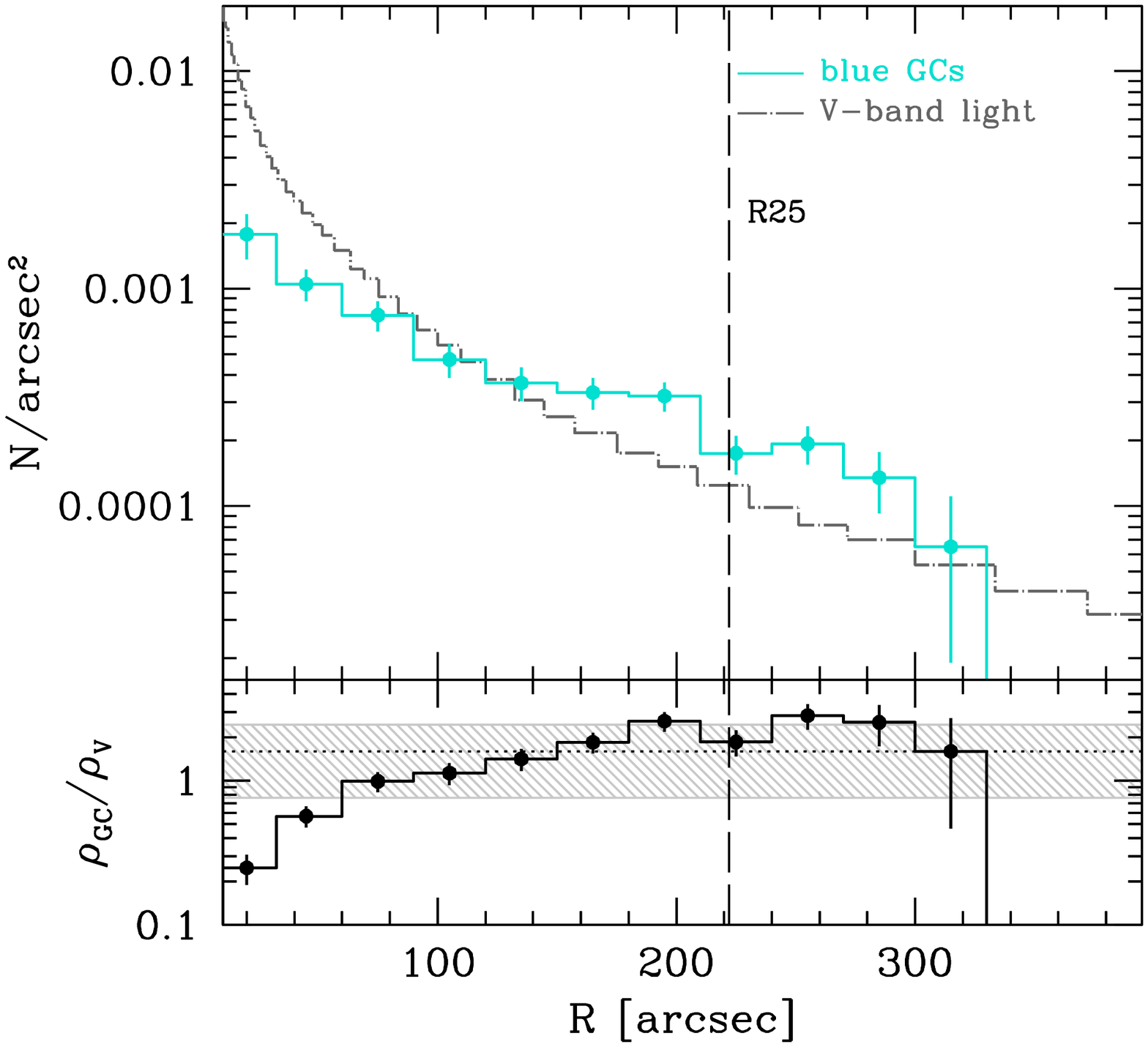}
}
\caption{Comparison between the radial distribution of red GCs ({\em left panel}, solid orange line) and blue GCs ({\em right panel}, solid cyan line) with the radial distribution of the $V$-band light (dashed grey line). To compare the distribution of GC with the $V$-band radial profile, we plot the complete GCs sample ($z<22.2\,\rm{mag}$, see \S\ref{sec:gc_lmxb_spatial}). Error bars are $1\sigma$ uncertainty assuming a Poisson distribution. The $V$-band radial profile, in both panels, was renormalized as in Fig.~\ref{fig:lmxb_profiles} to facilitate comparisons. Both GC profiles were re-normalized to the mean $\rho_{\rm{GCs}}/\rho_{\rm{V}}$ ratio for red GCs between $\sim$$50\arcsec$ and $\sim$$200\arcsec$, which is rather flat. In the bottom panel of each figure we plot the ratio between the number of GCs and the mean value of the $V$-band light at the given radial bin. The horizontal dotted line shows the average value of the $N_{\rm{GCs}}$-to-$V$-band light ratio within the radial bins covered with both GCs detections and $V$-band light. The shaded region indicates the $1\sigma$ standard deviation of the mean. The X-ray sources within the $D25$ region of the nearby spiral galaxy NGC~4647 were excluded and the area of circular annuli corrected accordingly. }
\label{fig:opt_gc_sb_profile}
\end{minipage}
\end{center}
\end{figure*}

\section{Radial distributions of X-ray sources and globular clusters}
\label{sec:lmxb_spatial}

To construct the radial distribution of X-ray sources and GCs we used the grid of circular annuli from \citet{2013ApJS..204...14L}, which is matched to the spatially-resolved completeness function $K(L)$ of the same authors (see their Fig.~3). The annuli are centered in the center of NGC~4649 and extend from a radius $R= 0.17\arcmin$ out to a radius $R=7.5\arcmin$ with constant linear spacing of $0.5\arcmin$.

\subsection{Globular clusters}
\label{sec:opt_gc_spatial}

The radial profiles of the red and blue GCs are presented in Fig.~\ref{fig:opt_gc_sb_profile}. Red GCs are more centrally concentrated than blue GCs and follow more closely the distribution of the stellar surface brightness, with the exception of a flattening at the inner radii ($R\lesssim 40\arcsec$). This is typical of GC populations \citep[see e.g.,][and references therein; and also Fig.~5 of Strader et al.~2012]{2006ARA&A..44..193B}.

We compared the radial distribution of GCs sources with the composite radial profile of the $V$-band light of \citet{2009ApJS..182..216K}. The $V-$band profile was constructed from many different data sources \citep[see Table 2 in][]{2009ApJS..182..216K} and is in Vega magnitudes. We plot the complete GCs sample having $z<22.2\,\rm{mag}$. The $\rho_{\rm{GCs}}/\rho_{\rm{V}}$ ratio curves (bottom panel of Fig.~\ref{fig:opt_gc_sb_profile}) show that the profile of red GCs seems to be fairly consistent, between $\sim40\arcsec$ and $\sim200\arcsec$, with the $V$-band light and therefore with the stellar mass of the host galaxy (see Sect.~\ref{sec:field_gc_comp} below), although the ratio over the entire range could also be consistent with a slow increasing gradient. By contrast, the distribution of blue GCs does not follow the $V$-band profile, except for radii between $30\arcsec$ and $120\arcsec$ with the adopted normalization.

\begin{figure*}
\begin{center}
\begin{minipage}{160mm}
\hbox
{
\includegraphics[width=77mm]{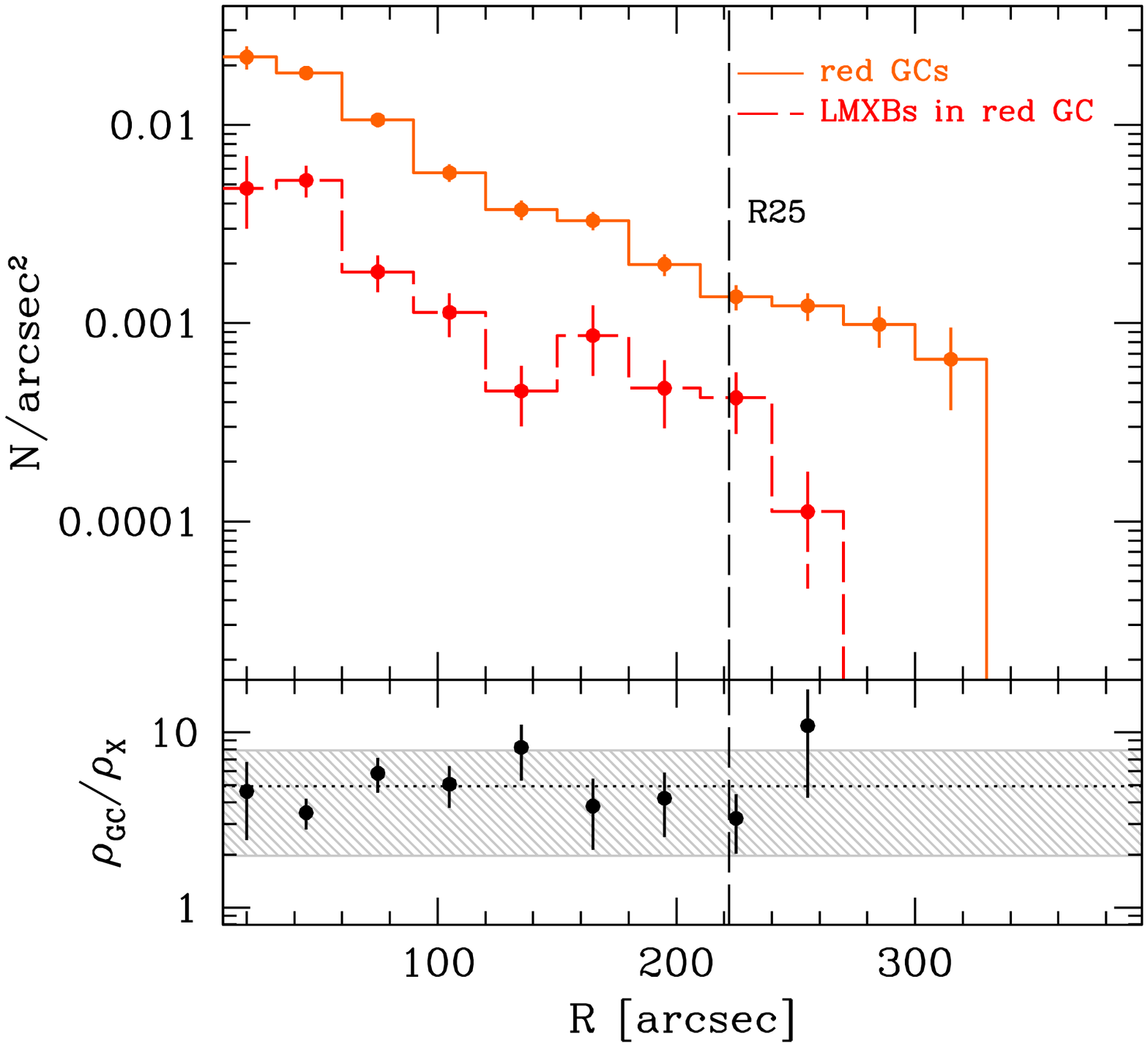}
\includegraphics[width=77mm]{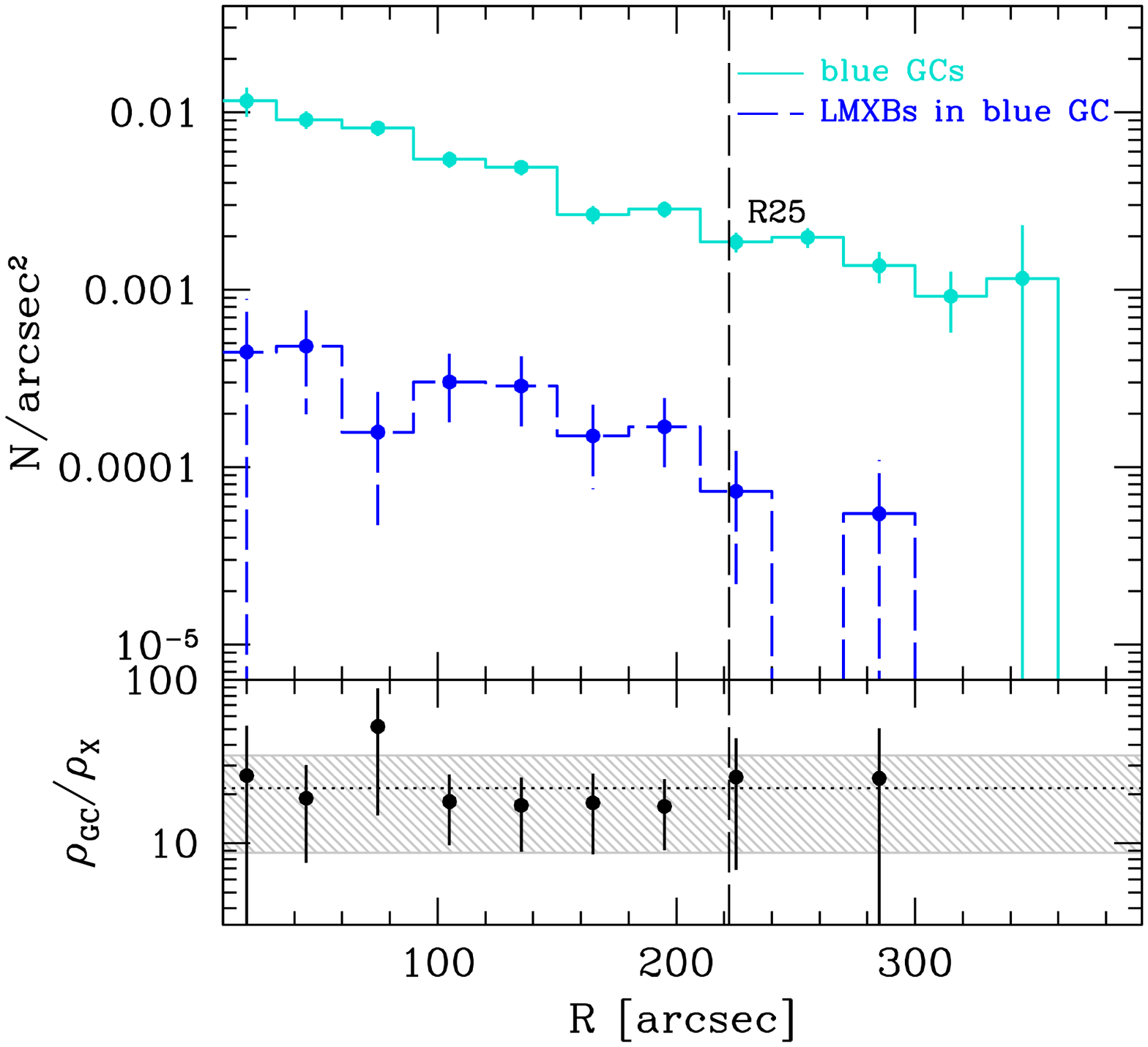}
}
\caption{Same as Figure \ref{fig:opt_gc_sb_profile}, but we did not apply a $z$-band magnitude cut.
Also plotted are the distributions of LMXBs detected in red (left, dashed red line) and blue (right, dashed blue line) GCs. Note that the effects of GC incompleteness in z-band are the same for the two (GCs and LMXBs) distributions. The X-ray incompleteness correction was also applied on the distributions of X-ray sources. 
In the bottom panel of each figure we plot the ratio between the number densities of GCs and X-ray sources ($\rho_{\rm{GCs}}/\rho_{\rm{X}}$) in each radial bin. The horizontal dotted line shows the mean value of the $\rho_{\rm{GCs}}$-to-$\rho_{\rm{X}}$ ratio within the first 10 radial bins, i.e. in the region where both the X-ray sources and GCs were detected. The shaded region indicates the $1\sigma$ standard deviation of the mean.}
\label{fig:gc_lmxb_vlight_comparisons}
\end{minipage}
\end{center}
\end{figure*}

\subsection{LMXBs in globular clusters}
\label{sec:gc_lmxb_spatial}

\begin{figure*}
\begin{center}
\begin{minipage}{160mm}
\hbox
{
\includegraphics[width=77mm]{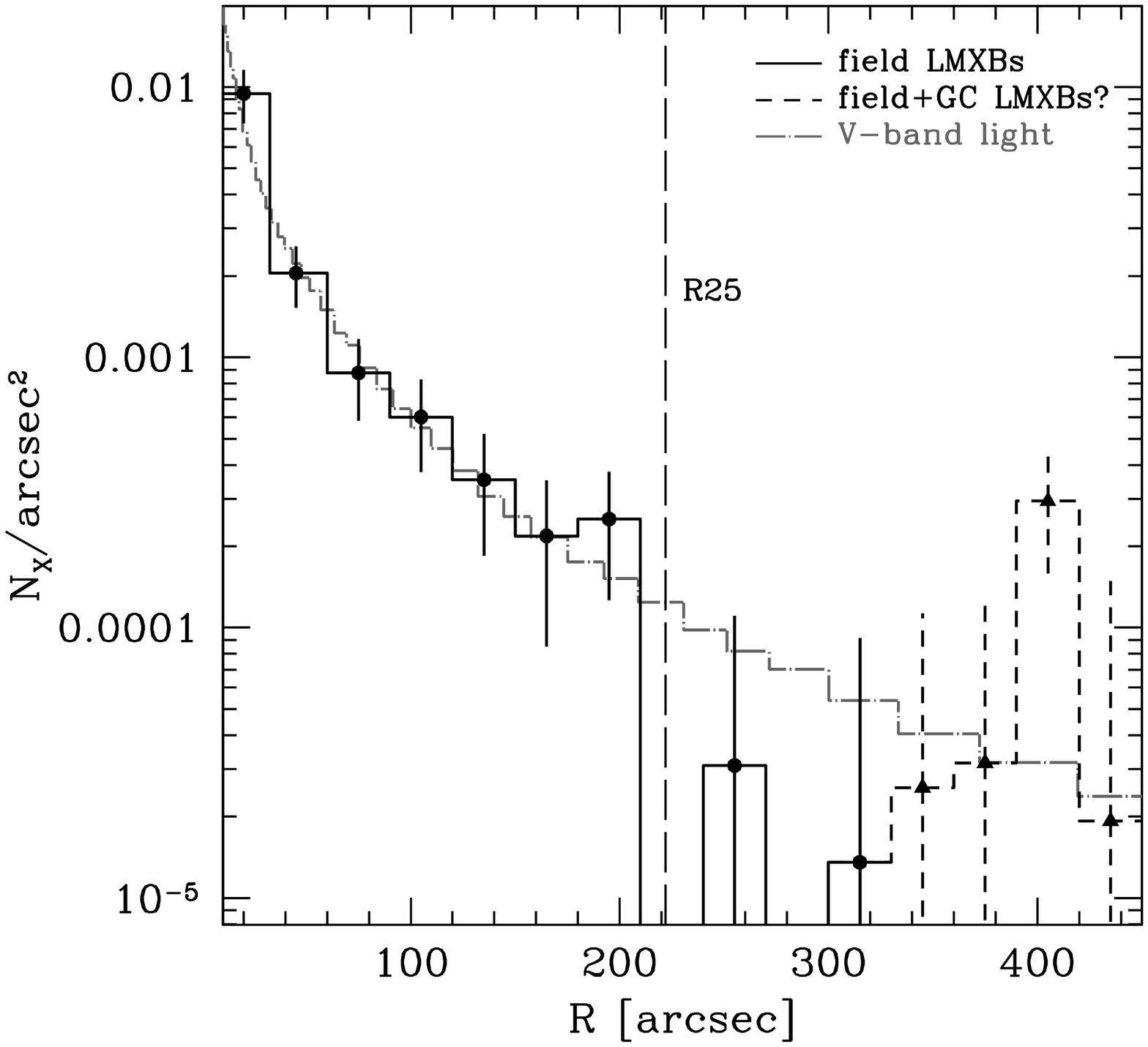}
\includegraphics[width=77mm]{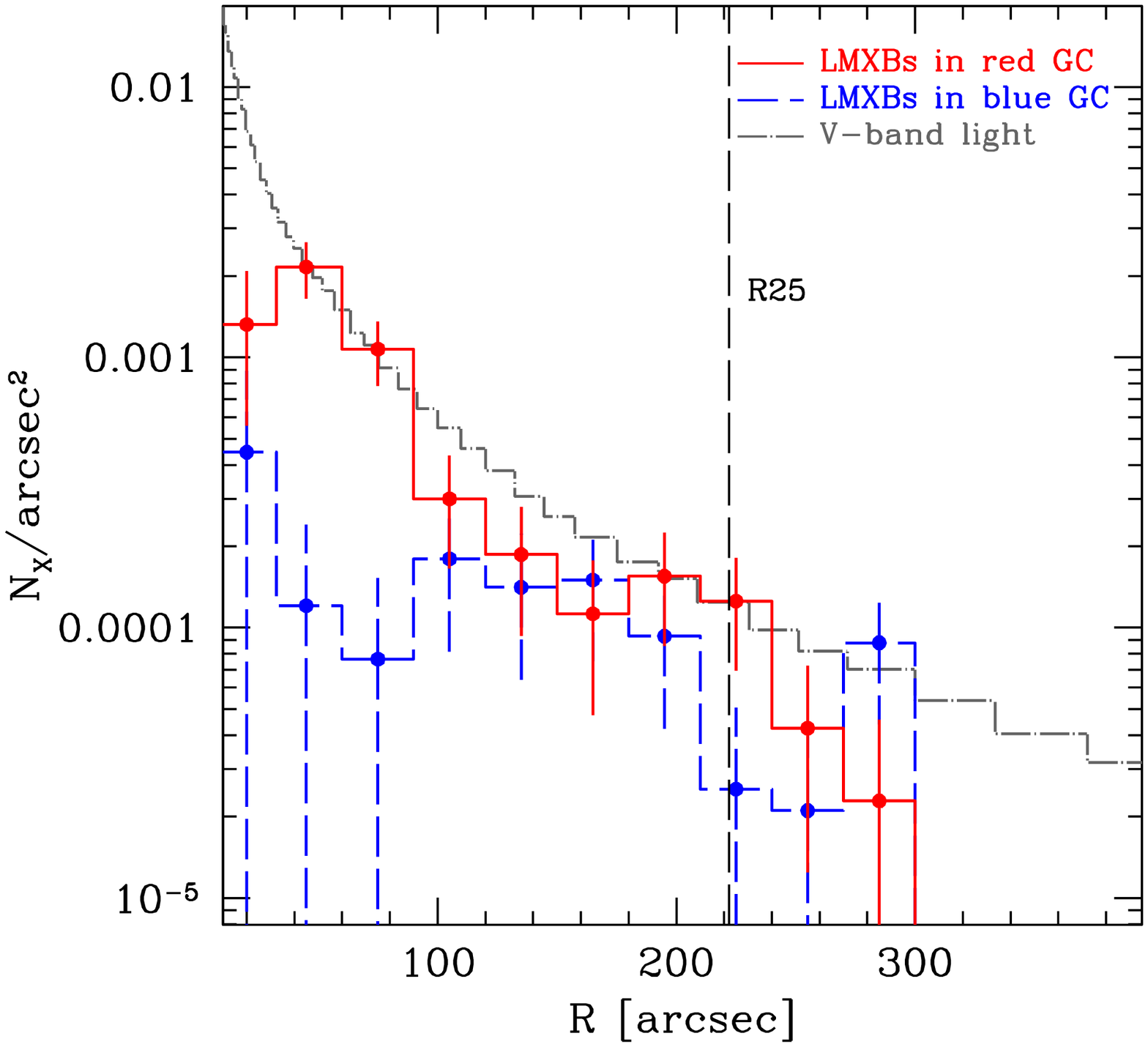}
}
\caption{Same as Figure \ref{fig:opt_gc_sb_profile}, but for the field LMXBs (left) and LMXBs in red/blue GCs (right) with both X-ray and optical incompleteness effects corrected (see \S\ref{sec:opt_gc_spatial} and \S\ref{sec:gc_lmxb_spatial} for details). For $R\gtrsim320\arcsec$ there is no optical data on GCs, therefore the field profile might be contaminated by GC-LMXBs and it is marked with a dashed line and black filled triangles.
The contribution of background AGNs per bin was subtracted only from the number of field-LMXBs. This correction was not necessary for the GC-LMXBs due to the GC association itself. The 20\% uncertainty on the background AGN number per bin is included in the error calculation. The $V$-band radial profile, in both panels, was renormalized in order to match the first bin of the profile for field LMXBs. The same normalization was used in Figure \ref{fig:opt_gc_sb_profile}.}
\label{fig:lmxb_profiles}
\end{minipage}
\end{center}
\end{figure*}

In Fig.~\ref{fig:gc_lmxb_vlight_comparisons} we compare the radial profiles of the red and blue GCs with the associated LMXBs. In this case, no $z$-band magnitude cut was applied to the GCs, to maximize the number of GC-LMXBs. The incompleteness correction was instead applied on the distributions of X-ray sources. In the bottom of each panel we show the ratio between the number density of GCs and GCs-associated X-ray sources ($\rho_{\rm{GCs}}/\rho_{\rm{X}}$) in each radial bin, along with their uncertainties. These ratios are uniform in the area of good {\it Chandra} and {\it HST} spatial coverage, meaning that on average the population of GCs with LMXBs has the same radial distribution as the parent red and blue GC distributions, as in NGC~1399 \citep{2011ApJ...736...90P}. A Kolmogorov-Smirnov (KS) test yields D values for red and blue GCs are $D_{\rm{red}}=0.25$, $D_{\rm{blue}}=0.28$, giving two-sided KS statistics probabilities of $0.91$ and $0.77$ respectively. This indicates that the distributions are likely drawn
from same underlying distributions. We have verified that the optical incompleteness does not affect the conclusion above, although it slightly changes the mean $\rho_{\rm{GCs}}/\rho_{\rm{X}}$ values since LMXBs are generally found in bright GCs. It also worsens the statistics of LMXBs at each radial bin, which results in larger error bars.

\subsection{LMXBs in the field}
\label{sec:field_lmxb_spatial}

To construct the radial distribution of field LMXBs we only used sources above the X-ray completeness limit and applied the X-ray incompleteness corrections. We calculated the area underlying the radial bins by taking into account the X-ray exposure map coverage as well as by subtracting the area corresponding to the $D25$ ellipse of the nearby spiral galaxy NGC~4647. Based on the corrected areas we subtracted the predicted contribution of background AGNs at each bin and converted the resulting number into number density. 

The resulting surface density of field LMXBs is compared to the $V$-band radial profile in the left panel of Fig.~\ref{fig:lmxb_profiles}. The uncertainties include the $1\sigma$ Poisson errors on the number of sources and both Poisson and 20\% uncertainty from cosmic variance on the expected AGN number per bin \citep[see Section 5.1 in][and references therein]{2012ApJ...749..130L}. We see clear agreement between the observed radial profile for field LMXBs and the $V$-band profile, within the $D25$ of NGC~4649; the KS test yields $D_{\rm{field}}=0.31$ and two-sided KS statistics probability of $0.56$. At $D25$ the data is suggestive of a possible dip. The face value of the statistical significance for the lack of sources, based on the $V$-band profile between $R25$ and $320\arcsec$, is $3.3\sigma$. However, this could be binning-related and should not be considered as a firm statistical conclusion. Using different binning we would lose information about the X-ray incompleteness, therefore we are not able to test the source dip further. Looking at both panels of Fig.~\ref{fig:lmxb_profiles} it is evident that the gap in the field-LMXB profile would be filled by the GC-LMXBs, suggesting that the lack of sources is not due to systematics. The KS statistics probability after we include the data outside the $D25$ ellipse, drops to $0.09$. At a radius of $\sim$$400\arcsec$ a moderately significant ($\approx2\sigma$) excess of LMXBs with respect to the $V$-light is observed. However, some of these ``field'' LMXBs may be associated with GCs, for which we lack optical coverage (Section~\ref{sec:data}). In Fig.~\ref{fig:lmxb_profiles} we mark these outer LMXBs with a different line.

\subsection{Comparison of LMXBs in GCs with field-LMXB}
\label{sec:field_gc_comp}

For a comparison of the radial distribution of GC-LMXBs with field-LMXBs, the incompleteness in both GC and X-ray samples was taken into account (Table~\ref{table:data}). To the GC sources we added the 12 matches with the GCs in the catalog of \citet{2008ApJ...674..857L} (see \S\ref{sec:data}).

The resulting radial distributions are shown in the right panel of Fig.\ref{fig:lmxb_profiles}, compared with the $V$-band light. As expected, LMXBs are preferentially found in red globular clusters, which also seem to follow the $V$-light profile, within statistics, except for the centermost bin. Considering only the {\it HST} sample, we find that red GCs contain 3.3 times more LMXBs than blue GCs (Table~\ref{table:data}) as already reported for this galaxy in \citet{2013ApJ...764...98K}.

\section{Discussion}
\label{sec:disc}

\subsection{LMXBs and stellar mass} 
\label{sec:lmxb_mstar_disc}

\citet{2004MNRAS.349..146G} showed that the number of LMXBs in old stellar populations scales with the stellar mass. \citet{2004ApJ...611..846K} reached the same conclusion, while also showing a dependence on the specific frequency of GCs  (i.e., number of GCs per unit mass) of each galaxy. The agreement between field LMXBs and $V$-band shows that this mass dependence of the LMXB population is uniform locally in NGC~4649, at least within $D25$.

\begin{figure*}
\begin{center}
\begin{minipage}{160mm}
\hbox
{
\includegraphics[width=77mm]{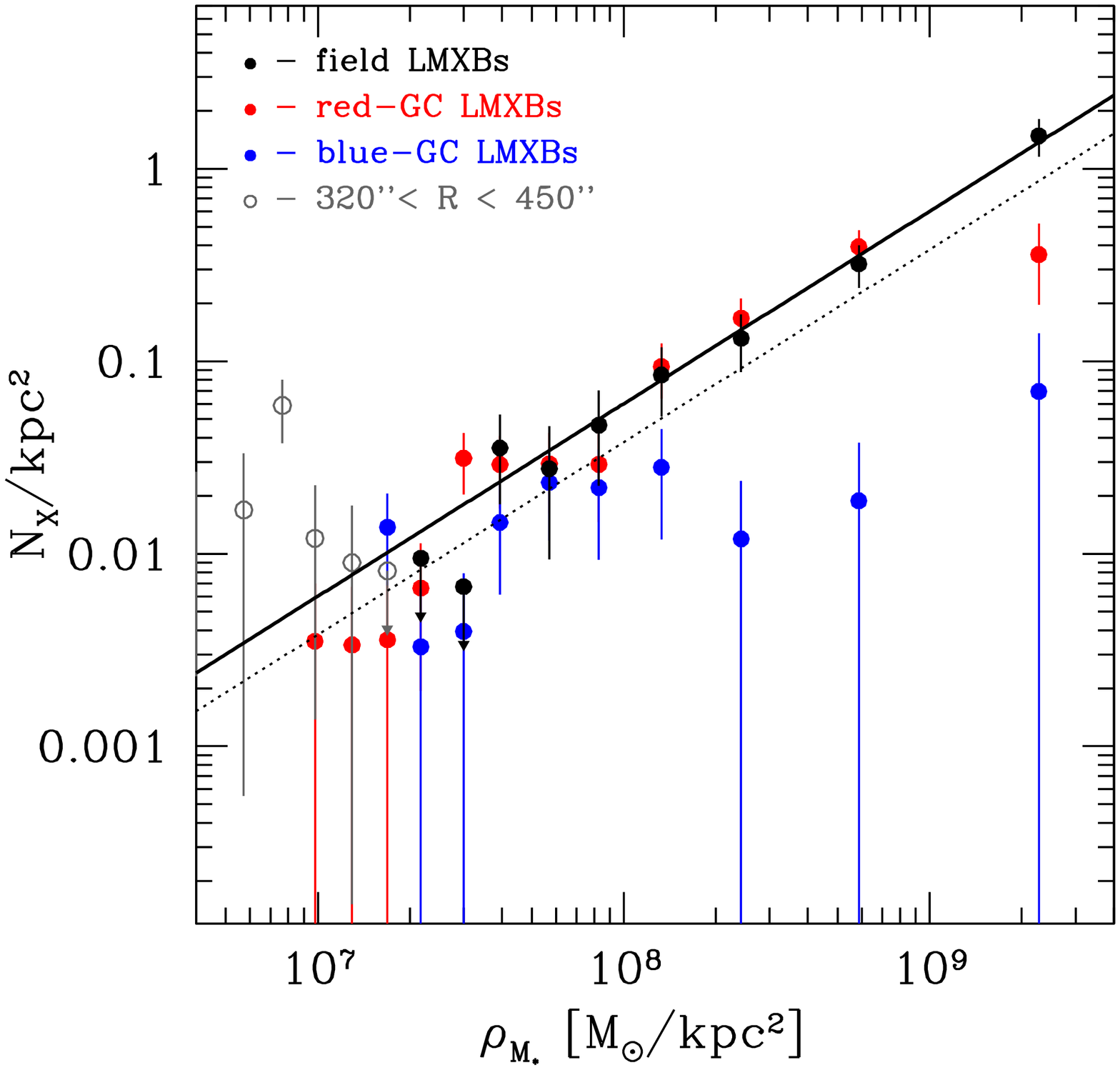}
\includegraphics[width=78mm]{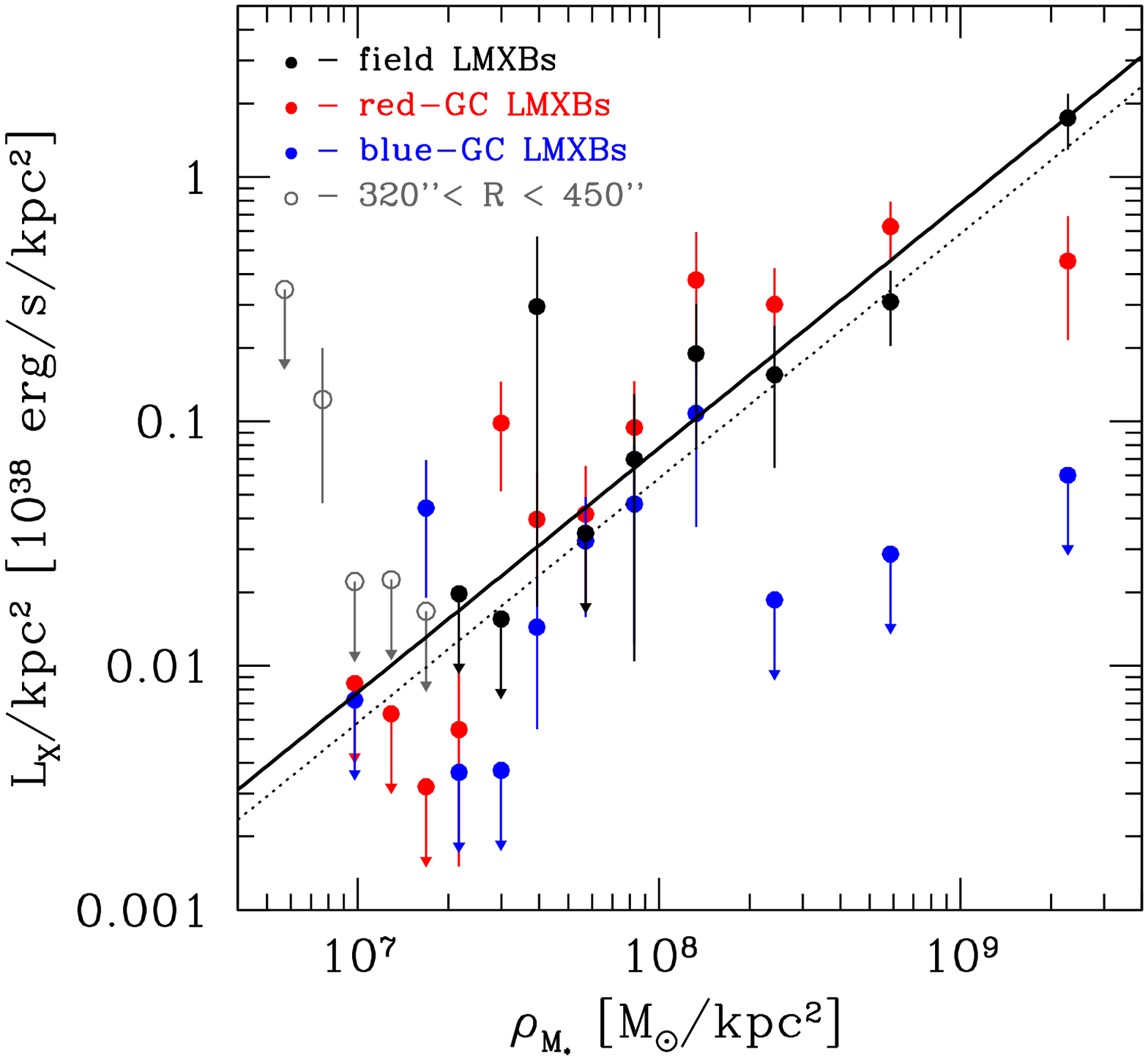}
}
\caption{Relation between the {\em local} stellar mass density in NGC~4649 ($M_{\odot}\,\rm{kpc}^{-2}$), the number density ($N_{\rm{X}}/\rm{kpc}^{2}$, {\em left} panel) and luminosity density ($L_{\rm{X}}/\rm{kpc}^{2}$, {\em right} panel) of LMXBs in the field (black filled circles) and in GCs (red and blue filled circles). The data were obtained using the $V$-band profile. Field sources at radii $R>320\arcsec$ may be contaminated by GC-LMXBs (see \S\ref{sec:disc}) and therefore marked with different symbols (grey empty circles). In all panels we plot the best-fitting linear relation (solid lines) for field LMXBs obtained using only the $V$-band profile and the galaxy-wide average relations (dotted lines) from \citet{2004MNRAS.349..146G}.}
\label{fig:nx_lx_mstar}
\end{minipage}
\end{center}
\end{figure*}

In Appendix \ref{sec:nx_lx_mstar_app} we derive a map of the stellar mass, which we compare with the $V$-band profile to obtain the mass conversion. The resulting number and luminosity density/mass plots are shown in Fig.~\ref{fig:nx_lx_mstar}. The best-fitting power-law, obtained via $\chi^{2}$ minimization for the number density/mass field LMXB sample is linear; the luminosity density/mass relation may be slightly flatter (Fig.~\ref{fig:nx_lx_mstar} and Table~\ref{table:fit_summary}). We excluded the points at $R>320\arcsec$ from these fits because of the lack of {\it HST} coverage (see Section~\ref{sec:data}). Within the uncertainties of the two relations, our best fit is consistent with \citet{2004MNRAS.349..146G}.

\begin{table*}
\tablewidth{0pt}
\tabletypesize{\scriptsize}
\centering
\begin{minipage}{115mm}
\caption{Summary of the parameters for $N_{\rm{X}}/\rm{kpc}^{2}-\rho_{M_{\star}}$, $L_{\rm{X}}/\rm{kpc}^{2}-\rho_{M_{\star}}$ relations obtained from $\chi^{2}$ fit.}
\label{table:fit_summary}
\begin{tabular}{@{}l l c c l c @{}}
\hline
\hline
\vspace{1mm}
& \multicolumn{3}{c}{{\sc free slope}} &  \multicolumn{2}{c}{{\sc linear fit}}\\
Relation &\vline\, $\log K$ & $\beta$ & $\chi^{2}/\rm{d.o.f.}$ & \vline\, $\log K$ & $\chi^{2}/\rm{d.o.f.}$\\
\hline
\multicolumn{6}{c}{{\sc field LMXBs}}\\
\hline
$N_{\rm{X}}/\rm{kpc}^{2}-\rho_{M_{\star}}$ &\vline\, $-9.29 \pm  0.80$ &  $1.01 \pm 0.09$ & $1.52/7$ &\vline\, $-9.22 \pm 0.05$ & $1.53/8$\\
$L_{\rm{X}}/\rm{kpc}^{2}-\rho_{M_{\star}}$  &\vline\, $-7.46 \pm  1.27$ &  $0.81 \pm 0.14$ & $6.80/5$ &\vline\, $-9.11 \pm 0.08$ & $8.49/6$\\
\hline
\vspace{0.1mm}
\end{tabular}
Note. The parameters are relative to the field LMXBs data fitted with power-law models $\log(N_{\rm{X}}/\rm{kpc}^{2}, L_{\rm{X}}/\rm{kpc}^{2}) = \log K + \beta\log(\rho_{M_{\star}})$, where $L_{\rm{X}}$ is in units of $10^{38}\,\rm{erg}\,\rm{s}^{-1}$, using the $\chi^{2}$ minimization technique, respectively setting the slope $\beta$ free and fixing it to unity. The error on both the slope and $K$ was computed with the standard $\Delta\chi^{2} = 1$ prescription. See Sect.~\ref{sec:disc} for details.
\medskip
\end{minipage}
\end{table*}

However, the \citet{2004MNRAS.349..146G} scaling relation was calibrated on the entire LMXB populations in nearby galaxies, with no separation between field and GC LMXBs. If we include GC-LMXBs, we obtain a 3.3 times larger scaling factor than \citet{2004MNRAS.349..146G}. This discrepancy may be due to the high globular cluster specific frequency of NGC~4649 ($S_{\rm{N}}=3.8 \pm 0.4$, Lee et al.~2008, $S_{\rm{N}}=5.16\pm1.20$, Peng et al.~2008), which is also important for the normalization of the LMXB luminosity functions \citep{2004ApJ...611..846K}. A further source of discrepancy between our calibration and that by \citet{2004MNRAS.349..146G}, may be related to the methods for mass estimates. \citet{2004MNRAS.349..146G} obtained the stellar mass for each galaxy through the calibration by \citet{2001ApJ...550..212B}. Our stellar mass is based on the calibration by \citet{2009MNRAS.400.1181Z}. According to \citet{2009MNRAS.400.1181Z}, there are minor systematic effects on mass estimates induced by different choices of IMF (Salpeter IMF in \citet{2001ApJ...550..212B}, \citet{2003ApJ...586L.133C} IMF in \citet{2009MNRAS.400.1181Z}) since the IMF can be considered uniform within a galaxy. More relevant discrepancies may be introduced by the assumption of other parameters, such as star formation history, metallicity, dust, which are expected to vary significantly from place to place within a galaxy.
Comparing our results with the $L_{\rm{X}}-M_{\star}$ relation from \citet{2004MNRAS.349..146G}, we found a larger normalization by a factor of $2.7$, consistent with the $N_{\rm{X}}-M_{\star}$ discrepancy.

\subsection{LMXB populations in red and blue GCs}
\label{sec:prop_gc}

The distribution of LMXBs in blue GC is uncorrelated with the stellar mass of the host galaxy and clearly differs from that of red-GC LMXBs, which follow the stellar mass distribution of the host galaxy except for the inner region ($R<30\arcsec$, $\rho\gtrsim 2\times 10^{9}\,M_{\odot}/\rm{kpc}^{2}$). Comparing the innermost bin of $\rho_{M_{\star}}$ in Fig.~\ref{fig:nx_lx_mstar} with the best-fitting $N_{\rm{X}}/\rm{kpc}^{2}-\rho_{M_{\star}}$ and $L_{\rm{X}}/\rm{kpc}^{2}-\rho_{M_{\star}}$ relations for field-LMXBs,  we find that both the number and luminosity densities of LMXBs in red GCs significantly deviate from the best-fitting linear model for field LMXBs (6$\sigma$ and 5.6$\sigma$ respectively). A similar deficit is seen comparing the red-GC population with the $V$-light. This deficit is well known and it is thought to be due to preferential destruction of GCs near the center. 

Fig.~\ref{fig:nx_lx_mstar} suggests also that in the region of $\rho_{M_{\star}}$ between $10^{8}$ and $10^{9}\,M_{\odot}/\rm{kpc}^{2}$, the distribution of red GC LMXBs in the $L_{\rm{X}}/\rm{kpc}^{2}-\rho_{M_{\star}}$ plane is skewed towards slightly higher luminosities if compared with field LMXBs. This difference cannot be due to the observed difference between the XLFs of GC and field LMXBs \citep{2009ApJ...703..829K}, because the latter becomes significant at $L_{\rm{X}}<5\times 10^{37}\,\rm{erg}\,\rm{s}^{-1}$, below our completeness limit.

We find 3.3 times more LMXBs in red than blue GCs (Table~\ref{table:data}). This ratio is typical for elliptical galaxies \citep[see e.g.,][]{2006ARA&A..44..323F} and in good agreement with previous results \citep{2003ApJ...595..743S, 2004ApJ...613..279J, 2006ApJ...647..276K, 2007ApJ...662..525K, 2007ApJ...660.1246S, 2009ApJ...703..829K}. \citet{2013ApJ...764...98K} demonstrated that this ratio is found at all LMXB luminosities. \citet{2006ApJ...636..979I} proposed that metallicity-dependent magnetic breaking with MS-donors may explain the observed difference, but this only works at lower luminosities ($< 2\times10^{37}\,\rm{erg}\,\rm{s}^{-1}$). To explain the metallicity effect of luminous NS-LMXBs, \citet{2012ApJ...760L..24I} further proposed that red giants serve as seeds for the dynamical production of bright LMXBs and the increase of the number densities and masses of red giants boosts the LMXB production for both WD and RG donors.

The number fraction of GC-LMXBs, i.e. the ratio of the number of GC-LMXBs (in both red and blue GCs) and the total number of LMXBs in both field and GCs, is $\sim$40\%, in agreement with the findings of \citet{2009ApJ...703..829K} based on deep {\it Chandra} observations of the three elliptical galaxies NGC~3379, NGC~4278 and NGC~4696. However, the total fraction of GC hosting a LMXB with $L_{\rm{X}} \gtrsim 4.8\times 10^{37}\,\rm{erg}\,\rm{s}^{-1}$, is $\sim$12\% (77 out of 626), a factor of $\sim$2 larger than that found by \citet{2009ApJ...703..829K}.  In particular, $\sim$18\% of red GCs (59 out of 329) and $\sim$6\% of blue GCs (18 out of 297) host a LMXB brighter than $4.8\times 10^{37}\,\rm{erg}\,\rm{s}^{-1}$. This inconsistency may be related to the extraordinarily luminous XLF of NGC~4649 \citep{2003ApJ...595..743S, 2004ApJ...600..729R}.

\subsection{GC specific frequency and LMXB formation}
\label{sec:spec_freq}
\begin{figure}
\begin{center}
\includegraphics[width=1\linewidth]{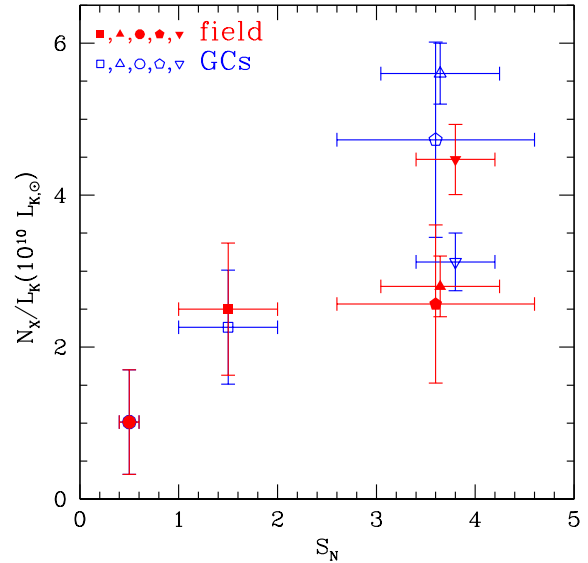}
\caption{Number of field-LMXBs (filled symbols) and GC-LMXBs  (open symbols) brighter than $4.8\times 10^{37}\,\rm{erg}\,\rm{s}^{-1}$ vs globular cluster specific frequency $S_{\rm{N}}$. The symbols represent: NGC~3379 (circles), NGC~4697 (squares), and NGC~4278 (pentagons) from \citet{2009ApJ...703..829K}. NGC~1399 (straight triangles) from \citet{2011ApJ...736...90P}, NGC~4649 (upside down triangles). For NGC~4946 we used the $S_{\rm{N}}$ estimate from \citet{2008ApJ...674..857L}, $3.8 \pm 0.4$.}
\label{fig:sn}
\end{center}
\end{figure}

\citet{2002ApJ...571L..23W} first noted a correlation between the abundance of LMXBs in early-type galaxies and the GC specific frequency $S_{\rm{N}}$, consistent with a part of the overall LMXB population at least being of GC origin. These results, however, could not address the question of the origin of the field LMXB: in situ or GC formation (e.g. Grindlay et al.~1984, Verbunt \& van den Heuvel~1995, Bildsten 
\& Deloye~2004). \citet{2005ApJ...631..511I} and \citet{2005ApJ...621L..25J} explored this relation to conclude that the field LMXB population is likely formed in situ. \citet{2009ApJ...703..829K} and \citet{2011ApJ...736...90P} pursued further these comparisons by using the GC and field LMXBs identified in four nearby galaxies. To these we can now add NGC~4649 (Fig.~\ref{fig:sn}).
Note that the values used in the present work are different than those of the original publications. For consistency, they are calculated using the same luminosity threshold as for NGC~4649 (our completeness limit $4.8\times 10^{37}\,\rm{erg}\,\rm{s}^{-1}$). For this galaxy we calculated the number of field- and GC-LMXBs within the $D25$ and normalized them for the $K$-band luminosity. In computing both the X-ray source numbers and the galaxy's $K$-band luminosity we excluded the central region as well as the $D25$ region of NGC~4647. We keep in mind that the higher $L_{\rm{X}}$ limit reduced more field-LMXBs than GC-LMXBs, because of the flattening of the GC-LMXB XLF \citep{2009ApJ...701..471V, 2009ApJ...703..829K}. We adopted the $S_{\rm{N}}$ estimate from the ground-based data by \citet{2008ApJ...674..857L}, $3.8 \pm 0.4$. 
We conclude that  overall the linear relation between $N_{\rm{X}}/L_{K}$ and $S_{\rm{N}}$ is stronger in GC-LMXBs, than in field-LMXBs, as suggested by \citet{2009ApJ...703..829K}, although the scatter is large. The presence of a (weaker) dependence of the field LMXB number density on $S_{\rm{N}}$ is consistent with a partial contribution of LMXBs originated in GCs to the native field LMXB population.

\subsection{Census of bright LMXBs in NGC~4649} 
\label{sec:ulx}

Three ULXs are detected in the field. This number is in agreement with the 2.4 sources predicted by the background AGN $\log N - \log S$ \citep{2008MNRAS.388.1205G} above $10^{39}$ erg s$^{-1}$.

One ULX is associated with a red GC (also in this case in agreement with the predicted number of background AGNs). The latter has $L_{\rm{X}}\approx2.14\times 10^{39}$ erg
s$^{-1}$ and is the source CXOU J124346.9+113234 identified and analyzed in depth by \citet{2012ApJ...760..135R}. These authors concluded that this object is a good candidate to be a BH radiating at super-Eddington luminosities, although they did not exclude the possibility that this is a highly super-Eddington NS. There is no evidence of ULXs in blue GCs. We searched for the source identification using SIMBAD\footnote{http://simbad.u-strasbg.fr/simbad/} astronomical database. We note that the counterparts are all within $0.8\arcsec$ from the detected X-ray source, which corresponds to the 99\% uncertainty circle of {\it Chandra} absolute positions for sources within $3\arcmin$ of the aimpoint.
Among the bright sources detected in the field, one is identified as X-ray source, one as X-ray binary and one as a Quasi-stellar object (Quasar) by  \citet{2011A&A...527A.126P, 2010MNRAS.406.1583W}. The first two sources are most likely ULXs. The results reported in the present section are summarized in Table~\ref{table:ulx}.

\begin{table}
\tablewidth{0pt}
\tabletypesize{\scriptsize}
\centering
\begin{minipage}{80mm}
\caption{Ultra-luminous X-ray sources ($L_{\rm{X}}\geq10^{39}\,\rm{erg}\,\rm{s}^{-1}$) in NGC~4649.}
\label{table:ulx}
\begin{tabular}{@{}c c c c c c@{}}
\hline
\hline
\vspace{1mm}
$\alpha_{J2000}$ & $\delta_{J2000}$ & $L_{\rm{X}}$ & Loc & ID & Separ\\
(h:m:s) & (d:m:s) & ($\rm{erg}\,\rm{s}^{-1}$) & & & (arcsec)\\
(1) & (2) & (3) & (4) & (5) & (6)\\
\hline
12:43:32.22 & 11:39:50.0 & $1.32\times 10^{39}$ & field & X & 0.8\\ 
12:43:36.53 & 11:30:09.4 & $5.5\times 10^{39}$ & field & XB & 0.4 \\
12:43:46.91 & 11:32:34.1 & $2.14\times 10^{39}$ & red GC & XB & 0.2\\
12:44:08.94 & 11:33:33.3 & $4.9\times 10^{39}$ & field & Q & 0.8\\
\hline
\vspace{0.1mm}
\end{tabular}
Note. (1) Right ascension (J2000); (2) declination (J2000); (3) 0.5--8 keV luminosity in $\rm{erg}\,\rm{s}^{-1}$;  (4) location (field/GC); (5) identification: X =  X-ray binary source, XB = X-ray binary, identifications by \citet{2011A&A...527A.126P}; Q= Quasi-stellar object, identification by \citet{2011A&A...527A.126P, 2010MNRAS.406.1583W};  (6) separation of the identified counterpart from the detected X-ray source. See \S\ref{sec:ulx} for details.
\medskip
\end{minipage}
\end{table}

\subsection{LMXBs in the galaxy's outskirts}
\label{sec:excess}
As discussed in Section~\ref{sec:field_lmxb_spatial}, the radial profile of field LMXBs departs from the $V$-light (stellar mass) profile outside the $D25$ of NGC~4649. It first suggests a relative lack of LMXBs and then it shows a possible excess at larger radii. The latter at face value would be consistent with similar peripheral overdensities reported by  \citet{2012arXiv1211.0399Z} in a sample of early-type galaxies, which included NGC~4649.  \citet{2012arXiv1211.0399Z} ascribed this flattening to both blue GCs and to formation kicks of sources connected with the stellar population at relatively smaller galacto-centric radii. In our case, the shape of the radial profile past $D25$, and in particular the 'dip', is hard to reconcile with the formation kick hypothesis, which would generate a general flattening of the source distribution. GC contamination at large radii of the field LMXB sample is possible because of the lack of deep GC data at those radii (see Section~\ref{sec:data}). The relative higher average luminosity of the field LMXB sample at these outer radii, suggested by the flatter than linear luminosity-density/mass distribution (Table~\ref{table:fit_summary}), would be consistent with this hypothesis, given the possible over-luminosity of GC-LMXBs relative tho those in the field (see Fig.~\ref{fig:nx_lx_mstar}, Section~\ref{sec:prop_gc}). 

An alternate possibility is that of a rejuvenation of the field LMXB population of NGC~4649 at large radii, caused by tidal encounters or accretion of companion or satellite galaxies. Such an interaction may be presently beginning with NGC~4647 \citep{2013ApJ...768...90L}. This scenario is also suggested by the anisotropy we find in the two-dimensional distribution of GCs and LMXBs in NGC~4649 at large radii (D'Abrusco et al 2013, in preparation). Population synthesis models of native field binary evolution suggest that the X-ray luminosity of these sources (per unit stellar mass) would evolve with age - younger sources would be relatively more luminous  \citep{2013ApJ...764...41F,2013arXiv1306.1405F}. This effect is not expected for GC sources \citep{2008MNRAS.386..553I, 2012arXiv1211.0399Z}

Taking the predictions by \citet{2013arXiv1306.1405F} at face value, an increase by a factor of $\sim$2.7 in $L_{\rm{X}}/M_{\star}$ compared to the value reported by \citet{2004MNRAS.349..146G} suggests that the younger population in the outskirts of NGC~4649 has a mean stellar age of approximately $4\pm1$ Gyr.

\section{Summary and conclusions}
\label{sec:soncl}
We have reported a large-area study of the LMXB and GC populations of NGC~4649. We find that:
\begin{inparaenum}[i\upshape)]

\item the LMXBs in the field follow the stellar mass distribution of the host galaxy within the $D25$ ellipse, but depart from it at larger radii. 

\item The LMXBs in red GC follow a radial distribution consistent with that of their parent red GCs. This distribution is consistent with the $V$-band light and therefore with the stellar mass of the host galaxy, except for the higher density region at $R<40\arcsec$, $\rho_{M_{\star}}\gtrsim 6\times10^{8}\,M_{\odot}/\rm{kpc}^{2}$, where they show a significant ($>3.5\sigma$) underdensity if compared with field LMXBs;

\item LMXBs in blue GC on the average have the same radial distribution as their parent blue GCs, which is uncorrelated with the stellar mass of the host galaxy. 

\item Comparing NGC~4649 with other four early-type galaxies for which similar quality data are available, we find that the LMXB population of NGC~4649 is consistent with a stronger dependence of the number of sources with the GC specific frequency for the GC-LMXBs than field LMXBs, although a weaker trend persists in the field LMXBs. 

\end{inparaenum}

Our results are consistent with a prevalent in situ origin for the field LMXBs, although with some contamination from LMXBs dinamically formed in red GCs. The deviations at outer radii of the LMXB radial profile from that of the stellar mass density may be in part due to contamination by GC-LMXBs, given our lack of GC coverage at these outer radii. The shape of the radial profile is not consistent with the hypothesis that these outer sources may be the results of formation kick displacement only \citep{2012A&A...546A..36Z}. We speculate that a rejuvenation of the field LMXB population \citep{2013ApJ...764...41F,2013arXiv1306.1405F} at larger radii, connected with tidal interactions and/or galaxy mergers, may be responsible for the excess high luminosity sources in these regions. 

\acknowledgments
SM is grateful to Stefano Zibetti for helpful discussions and data and to Marat Gilfanov for his valuable comments and suggestions to improve the quality of the paper. SM gratefully acknowledges financial support through the STFC grant 664 ST/K000861/1. TF acknowledges support from the CfA and the ITC prize fellowship programs. We thank Maurizio Paolillo for providing the data used in Fig.~\ref{fig:sn}. The authors thank the anonymous referee for helpful comments that improved this paper. This work was partially supported by NASA contract NAS8-03060 (CXC); NASA {\em Chandra} grants AR1-12008X and G01-12110X; and NASA {\em HST} grant G0-12369.01-A. We made use of \textit{Chandra} archival data and software provided by the \textit{Chandra} X-ray Center (CXC) in the application package CIAO. We also utilized the software tool SAOImage DS9, developed by Smithsonian Astrophysical Observatory. The FUV, and 24 $\mu$m images were taken from {\em Galex} and {\em Spitzer} archives, respectively. The \textit{Spitzer Space Telescope} is operated by the Jet Propulsion Laboratory, California Institute of Technology, under contract with the NASA. \textit{GALEX} is a NASA Small Explorer, launched in 2003 April. We also made use of data products from the Two Micron All Sky Survey (2MASS), which is a joint project of the University of Massachusetts and the Infrared Processing and Analysis Center/California Institute of Technology, funded by NASA and the National Science Foundation. Helpful information was found in the NASA/IPAC Extragalactic Database (NED) which is operated by the Jet Propulsion Laboratory, California Institute of Technology, under contract with the National Aeronautics and Space Administration. This research has made use of the SIMBAD database, operated at CDS, Strasbourg, France.


\appendix

\section{A. Stellar mass surface brightness map} 
\label{sec:mstar}

Stellar mass is the most important parameter in the investigation of gas-poor galaxies. In the last decade, several studies have proven the existence of a tight correlation between the collective number of LMXBs and the integrated stellar mass of the host early-type galaxies and bulges of late-type galaxies \citep[e.g.][]{2004ApJ...611..846K,2004MNRAS.349..146G}. 
This relation can now be explored in greater detail, using spatially resolved maps of stellar mass surface density which can be constructed based on optical and near-infrared (NIR) images, following the prescription from \citet{2009MNRAS.400.1181Z}. 

The fiducial method from \citet{2009MNRAS.400.1181Z} results from a careful comparison between stellar population synthesis models and multiband images. They use images from both Two Micron All Sky Survey (2MASS) and the Sloan Digital Sky Survey (SDSS) to express the stellar mass-to-light ratio $(M/L)$ as a function of the $H$-band ($1.66\,\mu\rm{m}$) and of the colors $(g-i)$ and $(i-H)$ ($g$- and $i$-bands having effective wavelength $475\,n\rm{m}$ and $763\,n\rm{m}$ respectively). The two sets of images have reasonably similar resolution ($2.5\arcsec$ and $1.3\arcsec$ FWHM, respectively). The resulting stellar mass surface density maps have typical accuracy of $<30\%$ at any pixel.
 
According to the same authors, in early-type galaxies, which lack young stellar populations, a good approximation of their method can be obtained using only one color, $(g-i)$. We adopted this recipe to construct the spatially-resolved stellar mass distribution in NGC~4649. Following the results of \citet{2009MNRAS.400.1181Z} summarized in their Table~B1, and based on the power-law fits to the $M/L$ as a function of one color, the stellar mass at each pixel $j$, in units of $M_{\odot}$, was estimated as follows:
\begin{equation}
\label{eq:stellar_mass}
\Sigma_{M_{\star},j} = \Sigma_{\rm{H},j} \Upsilon_{\rm{H},j}(g-i) 
\end{equation}
where $\Sigma_{\rm{H},j}$ is the $H$-band surface brightness and $\Upsilon_{\rm{H},j}(g-i)=10^{-1.222+0.780(g-i)_{j}}$.

We used background-subtracted images of NGC~4649 in $H$-band from 2MASS Large Galaxy Atlas\footnote{http://irsa.ipac.caltech.edu/applications/2MASS/LGA/} (LGA) and in $g$- and $i$-bands from the SDSS-III\footnote{http://www.sdss3.org/index.php} public archive. The SDSS images are mosaics of ``corrected frames", which are calibrated and sky-subtracted. As the pixel scales of the 2MASS and SDSS images are different ($1\arcsec/\rm{pix}$ and $0.396\arcsec/\rm{pix}$, respectively), we have degraded the SDSS $g$- and $i$-band images in order to match the resolution of the 2MASS $H$-band image. This was done using the routine {\sc hastrom}, from the NASA IDL Astronomy User's Library\footnote{http://idlastro.gsfc.nasa.gov/}, which properly interpolates without adding spatial information. 

Due to the typical low signal-to-noise ratio ($S/N$) of the SDSS images, we adaptively smoothed the $g$-band and $i$-band images before transforming them into maps of magnitude per pixel. We only applied slight smoothing in order to overcome the $S/N$ degradation at outer radii. We used {\sc adaptsmooth}\footnote{http://www.arcetri.astro.it/~zibetti/Software/ADAPTSMOOTH.html}, a code developed by \citet{2009MNRAS.400.1181Z} that enhances the $S/N$ with a minimum loss of effective resolution and by keeping the photometric fluxes unaltered. We ran {\sc adaptsmooth} for individual SDSS maps requiring a $S/N > 3$ per pixel and a maximum smoothing radius of 10 pixels. We set a background-dominated noise mode and used the background $rms$ in the two bands as input. The latter was measured by mean of the {\sc mmm} routine from the NASA's IDL Astronomy User's Library. Along with each smoothed image we obtained a mask that contains the smoothing radius for each pixel. The two masks were combined into a common mask having the maximum of the two smoothing radii at each pixel. In order to match the spatial resolution between the two bands, using the common mask as input we ran the adaptive smoothing again on the original $g$-band and $i$-band maps. We required the same $S/N$ threshold as before, but the smoothing radius at each position was provided in the input mask. The new $g$-band and $i$-band maps were converted into maps of magnitude per pixel in the respective bands. The ($g-i$) color map was obtained by subtracting at each pixel the $i$-mag from the $g$-mag.

The $H$-band image was left in its original resolution. Its pixel values were converted into luminosities in solar units, assuming solar $H$-band magnitude of $3.32\,\rm{mag}$ \citep{1998gaas.book.....B} and a distance to NGC~4649 of 16.5 Mpc. 

\begin{figure*}
\begin{center}
\includegraphics[width=1.0\linewidth]{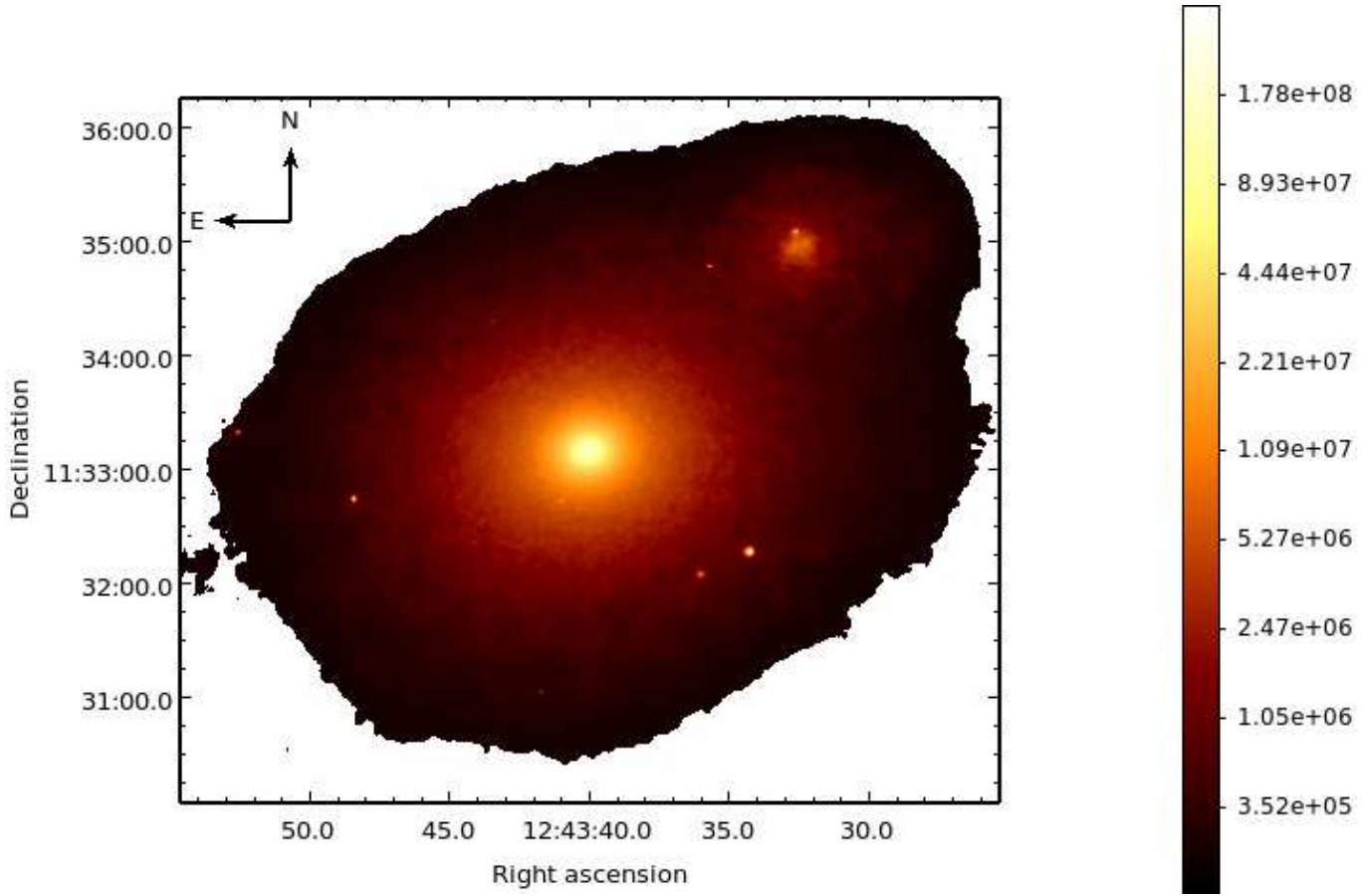}
\caption{Stellar mass surface brightness map for NGC~4649 and NGC~4647, in units of $M_{\odot}$, obtained by combining 2MASS $H$-band and SDSS $g$- and $i$-band images according the prescription from \citet{2009MNRAS.400.1181Z}. The map has the same pixel coordinates as the 2MASS $H$-band image. See Sect.~\ref{sec:mstar} for details. The spiral galaxy NGC~4647, approximately $2\arcmin.5$ away from NGC~4649, is also shown in the stellar mass image, although we excluded it from the analysis.}
\label{fig:mstar}
\end{center}
\end{figure*}

\begin{figure*}
\begin{center}
\includegraphics[width=1.0\linewidth]{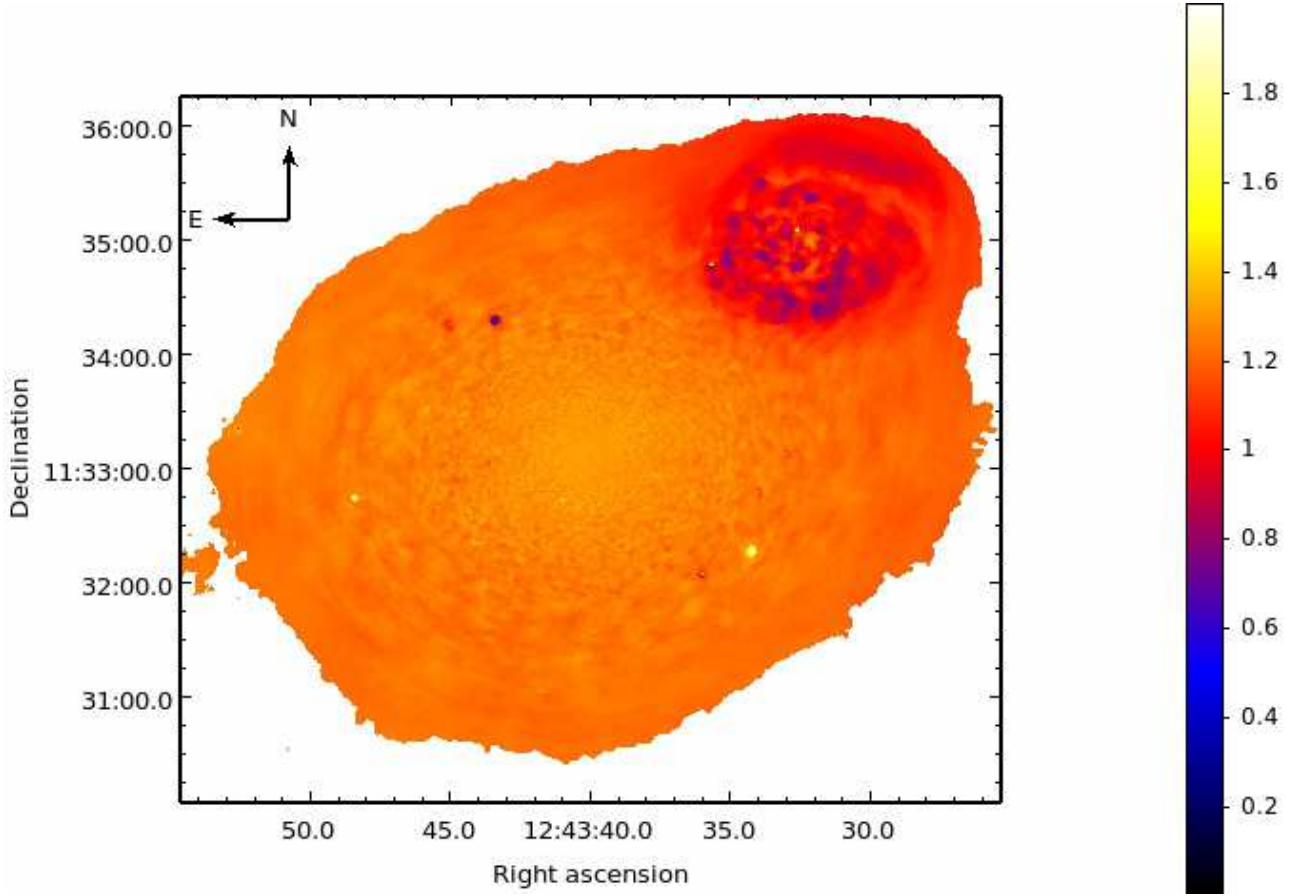}
\caption{Color map for NGC~4649 and NGC~4647, in units of $(g-i)$, obtained by combining SDSS $g$- and $i$-band images.}
\label{fig:color}
\end{center}
\end{figure*}

The maps of $H$-band surface brightness and $M/L$ were finally combined following eq.~(\ref{eq:stellar_mass}) to obtain the final image of stellar mass surface density with the same pixel coordinates as the 2MASS $H$-band image. A more elaborate version of this map, where heavier adaptive smoothing was applied for more qualitative purposes, is shown in Fig.~\ref{fig:mstar}. Fig.~\ref{fig:color} shows the color map in units of $(g-i)$, obtained by combining SDSS $g$- and $i$-band images. Note the difference between the young stellar population in NGC~4647 and the old in NGC~4649. 

\subsection{A.1. Comparison with NIR and Optical light distributions}
\label{sec:nir_opt_profiles}

\begin{figure*}
\begin{center}
\begin{minipage}{160mm}
\hbox
{
\includegraphics[width=0.5\linewidth]{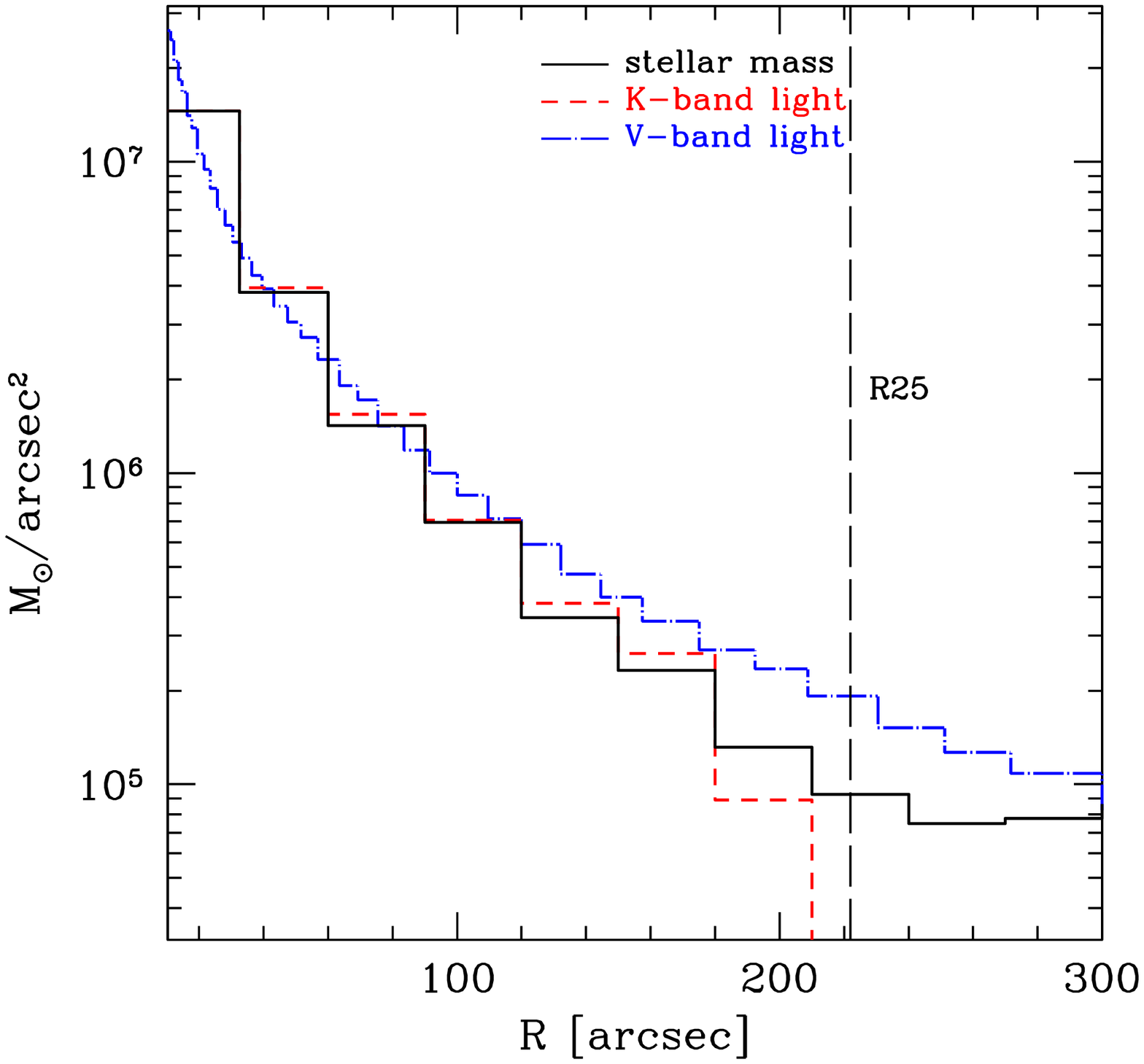}
\includegraphics[trim=1mm 55mm 5mm 5mm, width=0.52\linewidth]{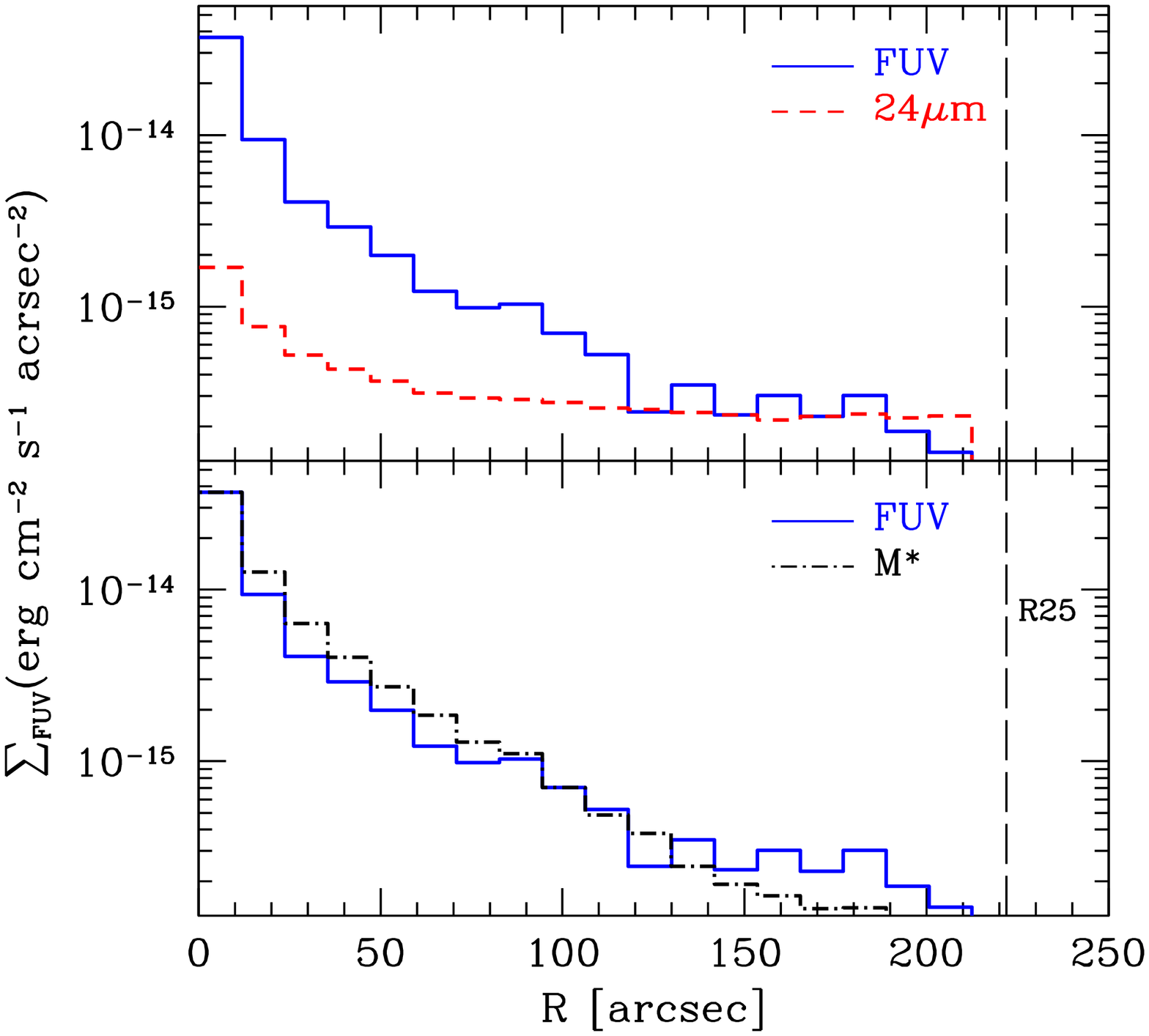}
}
\caption{{\em Left:} the radial profile of the stellar mass (solid, black line) within the $D25$ ellipse of NGC~4649, in units of $M_{\odot}/\rm{arcmin}^{2}$. The stellar mass distribution was obtained by combining 2MASS $H$-band and SDSS $g$- and $i$-band images (see Sect.~\ref{sec:mstar} for details). For comparison, we also plot the $K_{\rm{S}}$-band light (dashed, red line) from 2MASS archival data and the $V$-band radial profile (dash-dotted, blue line) from \citet{2009ApJS..182..216K}. These two curves are re-normalized in order to match the first bin of the stellar mass distribution. The semi-major axis of the $D25$ ellipse is shown with a dashed vertical line.{\em Right:} ({\em Upper panel:}) the radial distributions of FIR and FUV light in NGC~4649. For a qualitative comparison between the two, we renormalized the FIR radial profile in order to match the FUV profile at the bin of lowest surface brightness. The FIR distribution appears to be flatter and less extended than the FUV. ({\em Lower panel:}) the radial distributions of stellar light, $M_{\star}$ (see Appendix~\ref{sec:mstar}), and FUV light in NGC~4649. The stellar light  was renormalized in order to match the first bin of the FUV surface density. As expected \citep{1999ARA&A..37..603O}, the two profiles seem to match along the galaxy extent. The semi-major axis of the $D25$ ellipse is shown with a dashed vertical line.}
\label{fig:fuv_fir_profiles_gc_lmxb_vlight_comparisons}
\end{minipage}
\end{center}
\end{figure*}

We performed a qualitative check on the stellar mass distribution resulting from the map obtained in the previous section. We extracted the stellar mass radial profile using the same set of 15 concentric circular annuli defined in \citet{2013ApJS..204...14L} (see \S\ref{sec:lmxb_spatial} for details). The stellar mass radial profile within the $D25$ ellipse of NGC~4649, in units of $M_{\odot}/\rm{arcmin}^{2}$, is shown in Fig.~\ref{fig:fuv_fir_profiles_gc_lmxb_vlight_comparisons} (left panel, solid, black line). Using the same set of radial bins, we constructed the radial profile of the light in $K_{\rm{S}}$-band ($2.16\,\mu\rm{m}$). We re-normalized the latter in order to give the same value as the stellar mass profile in the first radial bin. We plot the radial profile of the $K_{\rm{S}}$-light in Fig.~\ref{fig:fuv_fir_profiles_gc_lmxb_vlight_comparisons} (left panel) with a red, dashed line and we note that it is consistent with the profile of stellar mass.

To check the effects of the $S/N$ degradation of 2MASS and SDSS images at outer radii on the stellar mass distribution, we plot in the same figure the surface brightness of the $V$-band magnitude (blue, dash-dotted line), provided by \citet{2009ApJS..182..216K} in a number of radial bins larger than that used for the stellar mass and $K_{\rm{S}}$ profiles. This profile is in units of mag/arcsec$^{2}$ and extends out to $R\sim$$11.5\arcmin$ from the center of NGC~4649. We converted it into units of Jy/arcsec$^{2}$ by assuming $F_{0} = 3640$ Jy as flux at $m_{V}=0$. The average of the values of the $V$-band profile within $0.17\arcmin-0.5\arcmin$ (first radial bin) was used to re-normalize it in order to match the stellar mass profile at the same bin. 

From the left panel of Fig.~\ref{fig:fuv_fir_profiles_gc_lmxb_vlight_comparisons} it is evident that the three profiles are in good agreement up to a radius of $\sim$150$\arcsec$. At larger radii both the stellar mass and the $K_{\rm{S}}$ distributions start descending more steeply than the $V$-band profile. This is likely due to the effects of the $S/N$ degradation of 2MASS and SDSS images. The difference between the stellar mass and the $K_{\rm{S}}$ profiles at $\sim$200$\arcsec$ is due to the slightly smoothed ($g-i$) color map included in the stellar mass profile. The agreement of the two profiles at smaller radii confirms that the $H$-band light is the main driver for the stellar mass distribution along the galaxy (see Sect.~\ref{sec:mstar}).

Based on the considerations above, the stellar mass surface brightness map will be used for further analysis (see e.g., \S \ref{sec:disc}) only within a radius of $\sim$150$\arcsec$ from the center of NGC~4649. At larger radii we will instead use the $V$-band radial profile (see \S \ref{sec:lmxb_spatial} and \S\ref{sec:disc}). We adopt the latter as standard profile at outer radii because it is a composite profile, carefully constructed from as many data as possible \citep[see][for details]{2009ApJS..182..216K}.

\subsection{A.2. Derivation of the spatially-resolved $N_{\rm{X}}-M_{\star}$, $L_{\rm{X}}-M_{\star}$ relations using the stellar mass map}
\label{sec:nx_lx_mstar_app}

We started from the stellar mass surface brightness image obtained as described in Sect.~\ref{sec:mstar}. We use the stellar mass density map within a radius of $\sim$150$\arcsec$ ($\sim$$2.5\arcmin$) from the center NGC~4649, where the stellar mass radial profile is in good agreement with the $V$-band radial profile (see \S~\ref{sec:nir_opt_profiles} for details). The few bright spots in the mass surface density map are most likely foreground stars and were masked out. We also masked the $D25$ region of NGC~4647 and the circle at the center of NGC~4649 with radius $R = 0.17\arcmin$. The stellar mass surface brightness image $\Sigma_{M_{\star}}$, constructed in units of $M_{\odot}$, was transformed into a stellar mass {\em density} map $\rho_{M_{\star}}$, in units of $M_{\odot}\,\rm{kpc}^{-2}$. 

A set of stellar mass density bins, with constant logarithmic spacing, was defined. The X-ray source number and their collective luminosity were assigned to each bin of stellar mass density according to the stellar mass density value at the position of the X-ray source. We selected sources whose X-ray luminosity exceeds $4.8\times 10^{37}\,\rm{erg}\,\rm{s}^{-1}$, the same threshold as used to construct the profiles of field and GC-associated LMXBs (see \S\ref{sec:lmxb_spatial}).
The number and luminosity of field and GC-LMXBs per stellar mass density bin were then corrected for the effects of incompleteness, following the same procedure as described in Sect.~\ref{sec:field_lmxb_spatial}. From the incompleteness-corrected number and luminosity of field LMXBs per bin of stellar mass density, we subtracted the predicted contribution of background AGNs at a given bin. We counted the number of pixels in each bin of stellar mass density and their cumulative area. Knowing the area, based on the $\log{N}-\log{S}$ function from \citet{2008MNRAS.388.1205G}, we calculated the predicted number of background AGNs, and their luminosity above the same threshold luminosity as mentioned beforehand. The resulting number and luminosity of field-LMXBs were divided by the total area in $\rm{kpc}^{2}$ in each bin. In constructing the analogous distributions for LMXBs associated with red and blue GCs we did not apply the correction for background AGNs, because of the GC-association itself. As we compare the distributions of LMXBs in red and blue GCs, the optical incompleteness was not taken into account as its effects are the same for the two distributions. The error bars in both number and luminosity densities at each bin of stellar mass density were computed by taking into account the Poisson noise in the incompleteness-corrected number of sources detected in the bin.

The pixel-by-pixel analysis of the stellar mass density map described above could not be performed at large radii ($R>150\arcsec$), due to noise in the SDSS images. As the SDSS-based stellar mass profile and $V$-band profile seem to agree well in the central area ($R<150\arcsec$) and do not show any features at large radii (see \S\ref{sec:mstar}), we derived the $V$-band mass-to light ratio and computed the mass density from the $V$-band profile in the outer regions. 

The final values of surface density of X-ray point sources ($N_{\rm{X}}/\rm{kpc}^{2}$) and luminosity ($L_{\rm{X}}/\rm{kpc}^{2}$, in units of $10^{38}$ erg s$^{-1}$ kpc$^{-2}$) are plotted against the value of the stellar mass surface density in Fig.~\ref{fig:nx_lx_mstar}. Along with the data, in Fig.~\ref{fig:nx_lx_mstar} we show the best-fitting linear $N_{\rm{X}}/\rm{kpc}^{2}-\rho_{M_{\star}}$ and $L_{\rm{X}}/\rm{kpc}^{2}-\rho_{M_{\star}}$ relations obtained below.

\section{B. Infrared and ultraviolet light distributions}
\label{sec:ir_uv_light_distr}

The far-infrared (FIR) and ultraviolet (UV) light from galaxies provide a wide range of information on the properties of their stellar populations and are observed, with different morphological properties, in both late-type and early-type galaxies. The FIR emission originates from dust grains heated by ionizing UV photons, which can be produced by the photospheres of both young and old stars.  

In this section we discuss the FIR and UV emission in both the elliptical galaxy NGC~4649 and the spiral galaxy NGC~4647. We use publicly available GALEX far-ultraviolet (FUV, 1529\,\AA) and near-ultraviolet (NUV, 2312\,\AA) background-subtracted images from the All Sky Surveys (AIS) program\footnote{http://galex.stsci.edu/GR4/?page=mastform} and {\it Spitzer} MIPS $24\,\mu\rm{m}$ Large Field image (``post Basic Calibrated Data" products)\footnote{http://irsa.ipac.caltech.edu/applications/Spitzer/Spitzer/}. The $24\,\mu\rm{m}$ background was measured in a region away from the galaxies and subtracted from the NGC~4649 and NGC~4647 emissions. 

We report the detection of a bright infrared point source in the MIPS $24\,\mu\rm{m}$ Field at $\sim36\arcsec$ N-E from the center of NGC~4649. We attempted to identify the source, bearing in mind that the PSF of MIPS $24\,\mu\rm{m}$ images is $\sim6.0\arcsec$ FWHM. We first searched for foreground sources using the Naval Observatory Merged Astrometric Dataset (NOMAD)\footnote{http://www.usno.navy.mil/USNO/astrometry/optical-IR-prod/nomad} \citep{2004AAS...205.4815Z}. This catalog is a merger of data from the Hipparcos, Tycho-2, UCAC-2 and USNO-B1 catalogs, supplemented by photometric information from the Two Micron All Sky Survey (2MASS) final release point source catalog. The closest source found is at $\sim26\arcsec$ S-W from the IR source. A search in the SIMBAD database\footnote{http://simbad.u-strasbg.fr/simbad/} reveals the presence of a red, $(g-z)\sim1.5$, globular cluster, $190.9202879+11.5441650$, reported by \citet{2009ApJS..180...54J}, at a distance of only $4.8\arcsec$ S-E from the IR source. However, as we are unable to safely determine the nature of the aforementioned bright IR source, we decided to mask it and exclude it from the analysis of the FIR emission from NGC~4649.


\subsection{B.1. FIR and FUV light distributions for NGC~4649}
\label{sec:fir_fuv_n4649}

The $24\,\mu\rm{m}$ emission from elliptical galaxies is thought to be produced by circumstellar outflows of dust-rich gas from old, mass-losing, red giant stars by illumination from their photospheres \citep[see e.g.,][and references therein]{2007ApJ...660.1215T,2008ApJ...672..244T}. This is confirmed by the evidence of a tight linear correlation between the luminosity at $24\,\mu\rm{m}$ and the $K_{\rm{S}}$-band luminosity in elliptical galaxies \citep[e.g.,][]{2009ApJ...707..890T}. 

Similarly, the FUV radiation in elliptical galaxies is smoothly distributed. It follows the optical light from old stellar populations. In particular, old,  metal-rich, low-mass stars which are in the helium-burning phase of the horizontal branch are thought to originate the FUV radiation \citep[see][for an extensive review]{1999ARA&A..37..603O}. This makes the FUV emission in early-type galaxies a tracer of the stellar mass loss. The contributions from active nuclei and from a minority of massive stars in a old population are not important in most cases. The latter can be due to recent star formation events in elliptical galaxies associated to massive cluster cooling flows.

The radial distributions of FIR and FUV light in NGC~4649 are shown in the upper-right panel of Fig.~\ref{fig:fuv_fir_profiles_gc_lmxb_vlight_comparisons}. For a qualitative comparison between the two, we renormalized the FIR radial profile in order to match the FUV profile at the bin of lowest surface brightness. The FIR distribution appears to be flatter than the FUV, suggesting that the hot horizontal branch stars may have lost nearly all of their envelope, therefore we may see only a residual $24\,\mu\rm{m}$ emission which is weaker than the FUV \citep{1999ARA&A..37..603O}.

In the lower-right panel of Fig.~\ref{fig:fuv_fir_profiles_gc_lmxb_vlight_comparisons} we compare the FUV profile with the stellar mass distribution, obtained in Section~\ref{sec:mstar}. We have renormalized the latter in order to match the first bin of the FUV surface density. The two profiles seem to be consistent along the galaxy extent, confirming the scenario mentioned above. For a more quantitative comparison, we used the correlation between the luminosity at  $24\,\mu\rm{m}$ and the $K_{\rm{S}}$-band luminosity from \citet{2009ApJ...707..890T}. We measured the integrated $24\,\mu\rm{m}$ luminosity within the $D25$ region of NGC~4649 (from which the contribution from NGC~4647 was removed) and obtained $L_{24\mu\rm{m}}=1.4\times 10^{42}\,\rm{erg}\,\rm{s}^{-1}$. Given this luminosity, eq.~(1) from  \citet{2009ApJ...707..890T} predicts a $K_{\rm{S}}$-band luminosity $L_{K_{\rm{S}}}/L_{\rm{K},\odot} = 1\times 10^{12}$. Using 2MASS $K_{\rm{S}}$-band data, assuming solar $H$-band magnitude of $3.32\,\rm{mag}$ \citep{1998gaas.book.....B},  $(B-V)=0.97$ \citep{1991trcb.book.....D}, and distance 16.5 Mpc, we measured within the same area an integrated $L_{K_{\rm{S}}}/L_{\rm{K},\odot} = 3\times 10^{11}$, i.e. a factor of $\sim$3 difference from the predicted $L_{K_{\rm{S}}}/L_{\rm{K},\odot}$, but still within the $rms$ of eq.~(1) from  \citet{2009ApJ...707..890T}. This is consistent with the hypothesis that the FIR emission may be originated from an old stellar population. It is also in agreement with the fact that the observed FIR profile is flatter than the FUV profile (see upper-right panel of Fig.~\ref{fig:fuv_fir_profiles_gc_lmxb_vlight_comparisons}).

\subsection{B.2. Star formation activity in NGC~4647}
\label{sec:sfr_n4647}

\begin{figure*}
\begin{center}
\includegraphics[width=1.0\linewidth]{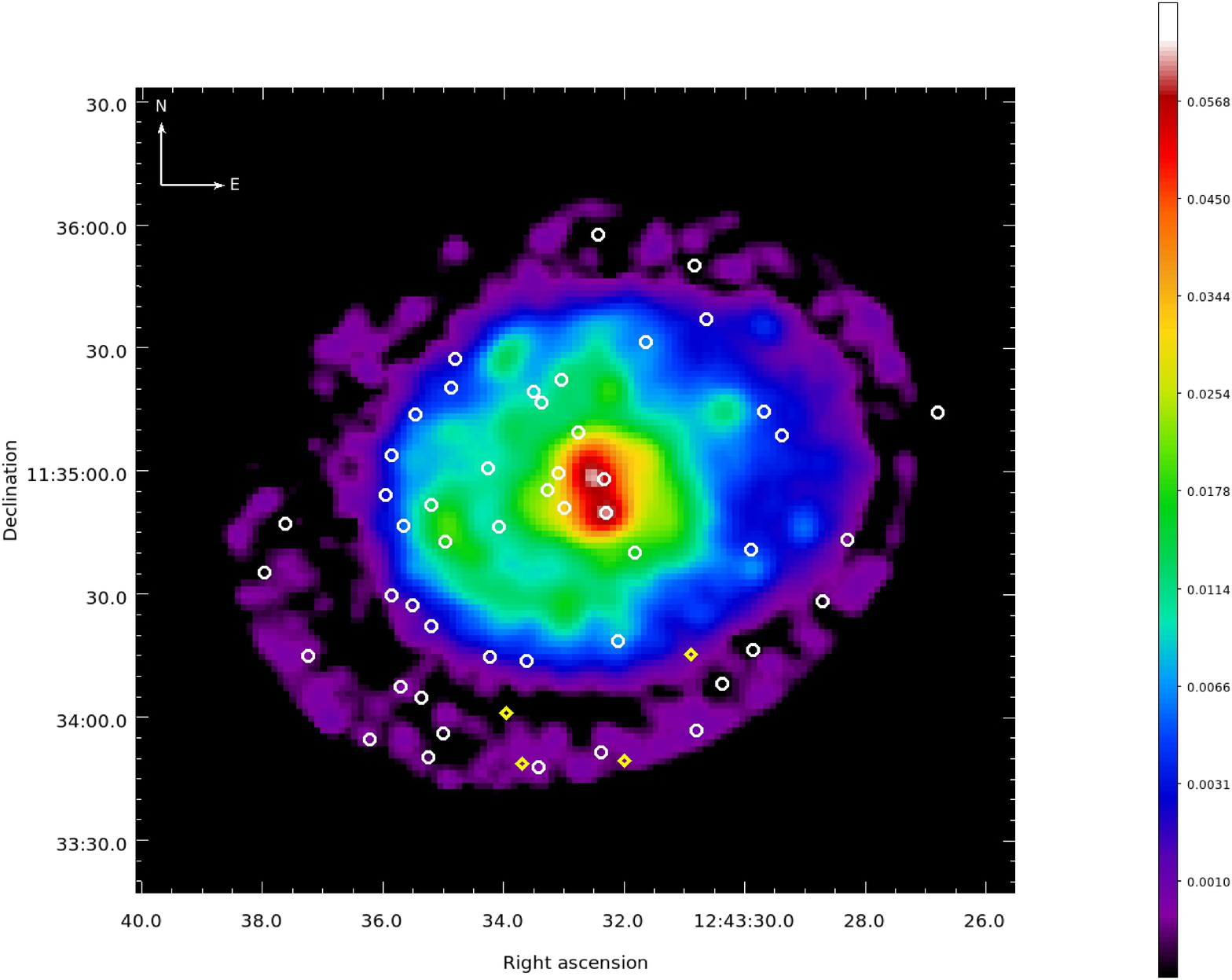}
\caption{Star formation rate density image for NGC~4647, in units of $M_{\odot}\,\rm{yr}^{-1}\,\rm{kpc}^{-2}$, obtained by combining GALEX FUV and {\it Spitzer} $24\,\mu\rm{m}$ images according the prescription of \citet{2008AJ....136.2782L}. See Sect.~\ref{sec:sfr_n4647} and \citet{2013ApJ...771..133M} for details on how the image was obtained. A slight smoothing was applied to the image, using a Gaussian kernel. The white circles and the yellow diamonds mark the locations of field-LMXBs and GC-associated LMXBs respectively, detected within the $D25$ ellipse NGC~4647.}
\label{fig:sfr}
\end{center}
\end{figure*}

The FIR and UV emissions from late-type galaxies have a substantially different origin than for early-types. The ultraviolet photons are irradiated from the photospheres of young, massive O and B stars, which are embedded in molecular clouds. As the latter are usually dust-rich, the dust grains are heated by absorbing the ionizing photons and emit FIR radiation. Both FIR and UV emission from spiral galaxies are distributed along the disk, tracing the location of the spiral arms and the star formation activity of the galaxy. 

The spiral galaxy NGC~4647 appears to be approximately $2\arcmin.5$ away from the elliptical NGC~4649. Due to their proximity, it is difficult to disentangle their X-ray populations. NGC~4647 should be dominated by short-lived ($<10-50$ Myr, $M_{opt} >5\,M_{\odot}$) high-mass X-ray binaries (HMXBs), in contrast to the long-lived ($> 1$ Gyr, $M_{opt} <1\,M_{\odot}$) low-mass X-ray binary (LMXB) population hosted by the elliptical galaxy NGC~4649. The number of HMXBs in late-type galaxies is proportional to the star formation rate (SFR) of the host galaxy \citep{2003MNRAS.339..793G,2012MNRAS.419.2095M}. In order to estimate the predicted number of HMXBs in NGC~4647, we used the calibration from \citet{2012MNRAS.419.2095M}. Accordingly, we calculated the integrated SFR inside the $D25$ ellipse of NGC~4647 using the combined near-ultraviolet (NUV) and FIR proxy from \citet{2006ApJS..164...38I}: 
\begin{equation}
\label{eq:sfr}
\rm{SFR}(M_{\odot}\,\rm{yr}^{-1}) = \rm{SFR}_{\rm{NUV,0}} + \rm{SFR}_{\rm{IR}}.
\end{equation}
The first term in eq.(A\ref{eq:stellar_mass}) is proportional to the near-ultraviolet (NUV) luminosity uncorrected for dust attenuation, taking into account the UV light from young stars escaping the dust clouds. The second term is instead based on the $8-1000\,\mu \rm{m}$ luminosity, tracing the emission from  dust surrounding the young stars \citep[see][and references therein for details, also about the conversion from $24\,\mu\rm{m}$ into $8-1000\,\mu \rm{m}$ luminosity.]{2012MNRAS.419.2095M}. The integrated SFR within the $D25$ ellipse of the spiral galaxy NGC~4647 is $\sim2.7\,M_{\odot}\,\rm{yr}^{-1}$. In the same region, \citet{2013ApJS..204...14L} detected 53 X-ray sources with luminosities greater than $9\times 10^{36}\,\rm{erg}\,\rm{s}^{-1}$. Of these sources, 49 are in the field and 4 are associated with GCs. Based on the $N_{\rm{X}}-\rm{SFR}$ in \citet{2012MNRAS.419.2095M} (their eq. (20)), the expected number of HMXBs in NGC~4647 having luminosity above the same threshold limit is $N_{\rm{X}}(L>9\times 10^{36}\,\rm{erg}\,\rm{s}^{-1})\sim 37$. 
Taking into account an accuracy of $rms = 0.34$ dex on this prediction, and considering that an unknown fraction of the 53 X-ray sources located within the $D25$ of NGC~4647 are LMXBs hosted by NGC~4646, the expected number of HMXBs in NGC~4647 is consistent with the observed number of X-ray binaries located within its $D25$. Therefore, in the following analysis we disregard the data within the $D25$ of NGC~4647.

To further investigate the star formation activity of the spiral galaxy NGC~4647 and its relation with the host X-ray binary population, we obtained the spatially-resolved map of SFR surface density, which is displayed in Fig.~\ref{fig:sfr}. This map was constructed from {\it Spitzer} and GALEX archival data, following the prescription by \citet{2008AJ....136.2782L} (their eq. D11):
\begin{equation}
\label{eq:sfr_density}
\Sigma_{\rm{SFR}} = 8.1\times 10^{-2} I_{FUV}+3.2\times 10^{-3} I_{24\,\mu m}.
\end{equation}
$I_{FUV}$ and $I_{24\,\mu m}$ are in units of $\rm{MJy}\,\rm{ster}^{-1}$. To combine the FUV and $24\,\mu\rm{m}$ images, which have reasonably similar angular resolutions ($4\arcsec$ and $6\arcsec$ FWHM, respectively) and sensitivities we followed \citet{2013ApJ...771..133M}, using the same tools as for obtaining the stellar mass map (see Appendix~\ref{sec:mstar} for details). After having spatially interpolated the $24\,\mu\rm{m}$ image in order to match the better resolution of the FUV image, we combined the two images and obtained the star formation rate density map for NGC~4647, in units of $M_{\odot}\,\rm{yr}^{-1}\,\rm{kpc}^{-2}$. From Fig.~\ref{fig:sfr} it is evident that most of the 53 X-ray sources detected within the $D25$ of NGC~4647 are located along the spiral arms of the host galaxy, confirming that they are actually dominated by HMXBs.

\end{document}